\documentclass[11pt]{article}
\usepackage{cite}
\usepackage{multicol}
\usepackage{graphicx}
\usepackage{amsmath}
\usepackage[a4paper]{geometry}
\usepackage{hyperref}
\usepackage{rotating}
\usepackage{xcolor}

\newcommand{\la}{\langle}
\newcommand{\ra}{\rangle}

\begin{document}

\begin{center}
  {\Large \bf The $\phi$--$\eta$ Puzzle and  the Five-Layer
    Effective-Temperature \vspace*{4mm}\\ Hierarchy
    in $pp$ Collisions at 2.76 and 7 TeV} 
\vskip1.0cm
M. Waqas$^{a}${\footnote{Corresponding author: (M. Waqas)
waqas\_phy313@yahoo.com, 20220073@huat.edu.cn}},
Wolfgang Bietenholz$^{b}${\footnote{wolbi@nucleares.unam.mx}}, and
Zhen-Yan Lu$^{c}${\footnote{luzhenyan@hnust.edu.cn}}
\\ \ \\
  {\small\it $^a$ Hubei Key Laboratory of Energy Storage and Power
    Battery \\ School of Optoelectronic Engineering, School of New
    Energy \\ Hubei University of Automotive Technology, Shiyan 442002,
    China \vspace*{1mm} \\
    $^b$ Instituto de Ciencias Nucleares, Universidad Nacional
    Aut\'{o}noma de M\'{e}xico \\  Apartado Postal 70-543,  04510
    Ciudad de M\'{e}xico, Mexico \vspace*{1mm} \\
    $^c$ Hunan Provincial Key Laboratory of Intelligent Sensors and
    Advanced Sensor Materials \\
    School of Physics and Electronics,
    Hunan University of Science and Technology \\
    Xiangtan 411201, China\\}
\end{center}

\noindent
The phenomenon of sequential kinetic freeze-out in high-energy
collisions has emerged as a crucial probe of the complex dynamics
governing the final stages of strongly interacting matter.
Inspired by the recently established quadruple freeze-out scenario
\cite{Badshah:2026ryg}, we present a thermodynamic analysis
of transverse momentum spectra for an extensive set of identified
particles, including non-strange, single- and multi-strange particles,
light nuclei, and the $\eta$ meson,
produced in proton-proton collisions at $\sqrt{s} = 2.76$ and $7$ TeV.
Using a modified Tsallis-Pareto function, we extract the
effective temperature and the non-extensivity parameter
for each species. Our analysis reveals a robust
five-layer hierarchy in the effective temperature.
At both energies, the layers are:
non-strange hadrons, singly-strange hadrons, multi-strange hadrons,
light nuclei, and the \(\phi(1020)\) meson. At
$7$~TeV, the \(\eta\) meson joins the \(\phi(1020)\) meson in the
highest-temperature layer. At both energies, each pair of neighboring
layers is separated by more than \(3\sigma\), and the separation
at the top is roughly \(7\sigma\) to \(8\sigma\). The ordering and
grouping remain stable at both collision energies, while the
temperatures shift upward at \(7\ \text{TeV}\). The
observed pattern demonstrates that the decoupling sequence is
governed by the interplay of strangeness,
cross-sections, and the OZI rule rather than by particle mass
alone. Notably, the \(\phi(1020)\) meson, despite having a mass
close to the proton, exhibits a significantly higher effective
temperature due to its OZI-suppressed coupling to the hadronic
medium. At 7 TeV,
the agreement between the temperatures of  \(\eta\) and \(\phi(1020)\)
holds within 5~MeV uncertainty.
This unexpected coincidence presents an intriguing puzzle that may
point to common early production mechanisms or in-medium effects
not yet fully understood. This establishes the five-layer hierarchy
as a robust feature, providing stringent new constraints
on theoretical models of hadronization.
These findings suggest that the sequential
decoupling of particles from the evolving medium is even more sensitive
to their internal structure and interaction history than previously
recognized, with important implications for our understanding of
QGP-like matter formation in small systems.\\


\newpage

\begin{multicols}{2}

{\section{Introduction}}
\label{intro}  
The exploration of strongly interacting matter under extreme
conditions is one of the most fascinating and rapidly evolving
areas of modern physics. When particles collide at ultra-relativistic
energies, whether in heavy-ion collisions {\it e.g.}\ at the Relativistic
Heavy Ion Collider (RHIC) or at the Large Hadron Collider (LHC),
or in high-multiplicity proton-proton ($pp$) collisions, the system
can reach temperatures and energy densities that mirror the
conditions of the early universe, just microseconds after the
Big Bang \cite{STAR:2005gfr,Busza}. Under such extreme conditions,
the theory of strong interactions, Quantum Chromodynamics (QCD),
predicts a transition from ordinary hadronic matter to a deconfined
state of quarks and gluons known as the Quark-Gluon Plasma (QGP)
\cite{Gyulassy:2004zyx,Gyulassy:2004zy,Ratti,Shuryak}. The discovery and
characterization of this exotic state of matter have been central
goals of high-energy nuclear physics for decades, offering unique
opportunities to study the fundamental constituents of matter and
the nature of the strong force \cite{Muller:2004kk}.

For many years, the formation of a QGP was primarily associated
with large collision systems such as gold-gold (Au-Au) or
lead-lead (Pb-Pb), where the large number of
participating nucleons and the resulting high energy density
were thought to be essential for deconfinement \cite{Bertulani:2015ana}.
Recent experimental findings, however, challenge this concept.
High-multiplicity $pp$ collisions
at the LHC have revealed striking signatures that resemble those
observed in heavy-ion collisions, including collective flow patterns,
enhanced strangeness production, and correlations that look like
the ridge phenomenon \cite{CMS:2015fgy,ALICE:2016fzo}. These
observations suggest that even small systems may create a medium
that exhibits a collective behavior, possibly indicating the formation
of a QGP-like state under extreme conditions \cite{Sahoo:2021aoy}.
This realization has promoted the study of $pp$ collisions from merely
a baseline for comparison to a vital testing ground for our
understanding of QCD matter in its most fundamental form.

The transverse momentum (\(p_T\)) spectra of identified particles
produced in these collisions serve as powerful probes of the
system's thermodynamic and transport properties \cite{STAR:2005gfr}.
As the fireball created in a high-energy collision expands and cools
down, it passes through several distinct stages of decoupling.

\begin{itemize}
  
\item The first
stage is chemical freeze-out, where inelastic collisions cease and
the relative abundances of different particle species stabilize
\cite{Braun-Munzinger:1995uec,Andronic:2005yp}. This stage is
typically studied through particle yield ratios and is characterized
by the chemical freeze-out temperature and baryon chemical potential.

\item
Following the chemical freeze-out, the system continues to evolve through
elastic collisions until the mean free path of particles becomes
comparable to the system size. At this point, even elastic interactions
stop, and the momentum distributions of particles freezes.
This stage is known as kinetic freeze-out, which is the last stage of
the system evolution. The parameters extracted from this stage,
namely the kinetic freeze-out temperature (\(T_0\)) and the average
transverse flow velocity (\(\langle \beta_T \rangle\)), provide
direct insight into the thermal and collective properties of the
expanding medium \cite{Schnedermann:1993ws}.

\end{itemize}

The effective temperature is another type of temperature which occurs
at the kinetic freeze-out stage or just before it \cite{Waqas:2020ioh}.
It includes the flow effect.

The study of kinetic freeze-out has a rich and evolving history,
leading to the development of distinct scenarios to describe
how different particle species decouple from the evolving medium.
The earliest and simplest hypothesis was the single freeze-out
scenario, which proposed that all particles, regardless of their
type or mass, decouple from the system simultaneously, sharing a
common set of freeze-out parameters \cite{Tang:2008ud,Broniowski:2001uk}.
In this framework, a single temperature and a single flow velocity
would describe the spectra of all particle species. While
appealing in its simplicity, this scenario was found to be inadequate
when confronted with high-precision experimental data, particularly
at LHC energies where the proton anomaly was observed
\cite{Rybczynski:2012ed}. This anomaly revealed the failure of
the single freeze-out picture to simultaneously describe the
transverse momentum spectra of pions, kaons, and protons.

The limitations of the single freeze-out scenario naturally led to the
proposal of the double freeze-out scenario. In this framework,
particles are grouped into two categories based on their strangeness
content: non-strange hadrons and strange hadrons. Each group is
then characterized by its own set of freeze-out parameters,
with strange hadrons decoupling earlier than non-strange hadrons due
to their smaller hadronic interaction cross-sections \cite{Shao:2009mu}.
The physical motivation for this scenario stems from the observation
that particles containing strange quarks interact less strongly with
the surrounding hadronic medium compared to their non-strange
counterparts. This reduced interaction strength means that strange
hadrons can decouple from the system while it is still relatively
hot and dense, whereas non-strange hadrons remain coupled for a
longer time \cite{Chatterjee:2014lfa}. 

A generalized approach, known as the mass-dependent or
multiple freeze-out scenario, emerged from the observation that
freeze-out parameters appear to vary continuously with the particle
mass rather than changing discretely between strangeness groups
\cite{Thakur:2016boy,Lao:2015zgd,Waqas:2022nae}. In this picture,
heavier particles decouple earlier because their
larger mass leads to a lower thermal velocity, which in turn reduces
their interaction rate in the expanding medium. Studies have shown
that the effective temperature extracted from transverse momentum
spectra increases systematically with increasing particle rest mass,
providing evidence for this mass-dependent differential freeze-out
picture \cite{Waqas:2020ioh,Lao:2015zgd,Wei:2016ihj,Waqas:2022fnl}.
This scenario represents the most general approach to understanding
freeze-out, as it does not impose any discrete groupings but rather
allows the freeze-out conditions to vary smoothly across the particle
spectrum. 

Our previous work \cite{Waqas:2021rmb} demonstrated the triple
freeze-out scenario, which represents a further refinement in our
understanding of the decoupling process. This framework distinguishes
between three distinct groups of particles: non-strange hadrons,
singly-strange hadrons, and multi-strange hadrons. Each of these
groups is assigned its own set of freeze-out parameters, reflecting
the different interaction strengths and production cross-sections
of these particle categories. The physical basis for this scenario
lies in the observation that the production cross-sections of
particles decrease as their strangeness content
increases.\footnote{Here and throughout this article, we use the
term ``strangeness'' in the sense of the absolute value,
keeping in mind that the strange quark has strangeness $-1$.}
Particles with larger cross-sections, such as pions and protons, remain
in equilibrium longer and freeze-out at lower temperatures. In contrast,
particles with smaller cross-sections, such as the multi-strange
\(\Xi\) and \(\Omega\) baryons, decouple earlier at higher temperatures. 

In a most recent study \cite{Badshah:2026ryg}, the quadruple
freeze-out scenario has been proposed to account for the freeze-out
behavior of composite particles, specifically light nuclei such as
deuterons, tritons, and helions. This scenario extends the triple
freeze-out framework by introducing a fourth distinct stage of
decoupling for these weakly bound composite objects. The quadruple
freeze-out framework proposes that non-strange hadrons, singly-strange
hadrons, multi-strange hadrons, and light nuclei each freeze-out at
a separate stage, characterized by a unique set of \(T_0\) and
\(\langle \beta_T \rangle\) values.
Ref.\ \cite{Badshah:2026ryg}
observed a clear hierarchy in the freeze-out temperatures,
with light nuclei exhibiting the highest freeze-out temperature,
followed by multi-strange hadrons, singly-strange hadrons, and finally
non-strange hadrons. This ordering is particularly intriguing because
it contradicts the naive expectation that loosely bound composite
objects would necessarily decouple at later stages when the system
is cooler and more dilute.

The natural progression of this work is to investigate whether
the freeze-out hierarchy can be extended even further. This
raises a compelling and important question: is the quadruple
freeze-out scenario the final word on the complexity of
freeze-out, or does a more detailed investigation, using a wider
range of particle species, reveal an even more intricate structure?
The present study explores this possibility
and tests the limits of the current understanding of freeze-out
dynamics. By performing a systematic analysis of transverse momentum
spectra for an extensive set of particles produced in $pp$
collisions at the LHC, measured by the ALICE experiment,
this work aims to determine whether a more
complex, multi-stage effective-temperature hierarchy provides a
superior description of the experimental data.
Using an extensive set of identified particles,
we find evidence for an additional freeze-out stage at the earliest
times, which cannot be accounted for by strangeness content or mass
alone. The full hierarchy and its physical interpretation are presented
in Section \ref{results}.

The choice to focus on $pp$ collisions is particularly
significant. While $pp$ collisions were traditionally regarded as
simple reference systems for understanding heavy-ion collisions,
the discovery of QGP-like signatures in high-multiplicity $pp$ events
has transformed their role
\cite{ALICE:2016fzo,Kalaydzhyan:2015xba,Bozek:2009dt}. Studying
the freeze-out in $pp$ collisions allows us to investigate the
fundamental mechanisms of particle production and collective
behavior in systems with a much smaller initial volume and
shorter lifetime compared to heavy-ion collisions. If an extended
freeze-out scenario is observed in $pp$ collisions, it would
suggest that the sequential decoupling of particles is a robust
feature of this collision system, motivating further studies
of its possible persistence in other systems. This could provide
deep insights into the nature of hadronization and the final-state
interactions in small systems. Furthermore, such a discovery could
have important implications for our understanding of the onset of
collectivity and the possible formation of a QGP-like medium in
high-multiplicity $pp$ collisions. The present analysis, and the
quintuple freeze-out scenario established in this work, are
restricted to $pp$ collisions at $\sqrt{s}=2.76$ and 7~TeV.
The two energies provide a consistency check of the observed
hierarchy.

The impact of this investigation extends beyond the mere
identification of a new freeze-out scenario. The existence of
an extended freeze-out would imply that the decoupling of
particles from the medium is even more sensitive to their
internal structure, binding energy, and interaction history than
previously assumed. This would have significant implications
for our understanding of the hadronization process and the nature
of the final state interactions in high-energy collisions.
Furthermore, the observation of additional freeze-out stages would
provide stringent new constraints on theoretical models of the QGP
and its evolution, potentially revealing new aspects of the complex
dynamics that govern the final stages of a nuclear collision.

Section \ref{method} describes the theoretical formalism and the
methodology employed in this analysis. Section \ref{results} presents
the results and discusses their physical implications. Finally,
Section \ref{conclu} summarizes the main findings
and presents our conclusions.

\section{The method and formalism}
\label{method}

The transverse momentum spectra of identified particles were
analyzed using the modified Tsallis-Pareto function (TP-like
function) \cite{Yang:2019rje,Tai:2020fkf}. This function is based
on Tsallis statistics
\cite{Ahmed:2026dhc,Tsallis:1987eu,Saraswat:2026xat,Nedorezov:2024bve,Parvan:2019bga,Parvan:2016mbv,Cleymans:2011in} and is well suited for
describing spectra over a wide \(p_T\) range, including both
the soft and hard regions. The particles in the high \(p_T\)
region are usually described by power-law distributions
\cite{ALICE:2010syw,Wilk:1999dr,PHENIX:2011rvu,ALICE:2013txf},
which are frequently used in high-energy physics.
The TP-like function \cite{Yang:2019rje,Tai:2020fkf} is given as
\begin{eqnarray}
\frac{d^2N}{dp_Tdy} &=& \frac{dN}{dy}C_0p_Tm_T\bigg(1+\frac{m_T-m_0}{n_0T}
\bigg)^{-n_0}, \nonumber \\
C_0 &=& \frac{(n_0-1)(n_0-2)}{n_0 T [n_0 T + (n_0-2)m_0]} , \nonumber \\
m_T &=& \sqrt{p_T^2 + m_0^2} \ ,
\label{eq1}
\end{eqnarray}
where $m_{T}$ is the transverse mass, \(m_0\) is
the rest mass of the particle, \(T\) is the effective temperature, and
\(n_0 > 0\) is the power index that describes the degree of non-equilibrium.
It is related to the non-extensivity parameter
\(q\) by \cite{Zhang:2021npy,Yang:2021bab}
\begin{align} \label{eq2}
n_0 = \frac{1}{q - 1} , \qquad q > 1 .
\end{align}  
A small \(n_0\) corresponds to a large \(q\), indicating that
the system is far from equilibrium, while a large \(n_0\)
corresponds to a small \(q\), indicating that the system is
close to equilibrium.

In terms of the probability density function of $p_T$, Eq.\ (\ref{eq1})
can be simplified to \cite{Cleymans:2012ya,Yang:2021bab}
\begin{equation} \label{eq3}
f(p_T) = C_0 \, p_T \, m_T \left[ 1 + \frac{m_T - m_0}{n_0 T} \right]^{-n_0},
\end{equation}
In order to add flexibility to the fits,
the function is modified by introducing
a correction index \(a_0\) \cite{Yang:2019rje,Tai:2020fkf},
\begin{equation} \label{eq4}
f(p_T) = C_0 \, (p_T m_T)^{a_0} \left[ 1 + \frac{m_T - m_0}{n_0 T} \right]^{-n_0},
\end{equation}
where \(a_0\) is a dimensionless parameter. When \(a_0 = 1\), the
modified function reduces to the standard Tsallis distribution.
We include \(a_0\) as a free parameter to
improve the fit quality, specifically at the very low
\(p_T\) $<$ 0.4 or 0.5 GeV/$c$. The normalization constant \(C_0\)
is treated as a free fit parameter and absorbs the overall
normalization and scale of the distribution. The dimension of
\(C_0\) is adjusted to $a_{0}$, such that the dimension of $f(p_T)$
remains invariant.

The TP-like function (\ref{eq4}) is fitted to the measured \(p_T\)
spectra of various particle species produced in proton-proton
collisions at \(\sqrt{s} = 2.76\) and \(7\) TeV. The fit parameters
\(T\), \(n_0\), and \(a_0\) were extracted for each particle species.
The effective temperature \(T\) was used to characterize the
freeze-out hierarchy. We emphasize that the effective temperature
\(T\) includes the effect of collective flow and is therefore not
identical to the kinetic freeze-out temperature \(T_0\) extracted
from blast-wave analyses.

In the present work, we use a single component
distribution. A multi-component distribution
can also be used to improve the fit if the single component does
not describe the high \(p_T\) region well. The high \(p_T\) region
is described by an inverse power law
\cite{Odorico:1982eq,Biyajima:2016fpb,UA1:1982fux}, {\it i.e.}\
the Hagedorn function \cite{Hagedorn:1983wk,ALICE:2014zxz}
\begin{equation} \label{eq5}
  f(p_T) = \frac{1}{N}\frac{dN}{dp_T}
  = X p_T \left(1 + \frac{p_T}{p_0}\right)^{-b} .
\end{equation} 
In this equation, $p_0$ and $b$ are the free parameters, while $X$
is the normalization constant. For the description of a wide \(p_T\)
region, two methods can be used: (I) the superposition principle
and (II) the usual step function. For details of these methods,
we refer to our previous work in Ref.\ \cite{Waqas:2022fnl}.

\section{Results and discussion}
\label{results}

We first discuss our results at $\sqrt{s} = 2.76~{\rm TeV}$.
Figure \ref{fig1-b} shows the \(p_T\) spectra of a variety
of particle species; it
consists of twelve different panels labeled from (A) to (L).
Panel (A) presents the \(p_T\) spectra of $\pi^+ +\pi^-$, while
the \(p_T\) spectra of $K^++K^-$, $p+\bar p$, $K^0_S$, $\Lambda$
and $\bar \Lambda$ are shown in panel (B) to (F), respectively.
Panels (G) to (L) show the \(p_T\) spectra of $\Xi^-$, $\bar \Xi^+$,
$\Omega^-+\bar \Omega^+$, $\phi$, deuteron $d$ and anti-deuteron $\bar d$,
respectively. The experimental data are represented by the symbols,
they are taken from Refs.\
\cite{ALICE:2014juv,Schuchmann:2015lay,Colella:2013iky,ALICE:2015cql}.
The curves over the experimental data are the fits to the function
$f(p_{T})$ in Eq.\ (\ref{eq4}). The model describes the data well. 
The values of the extracted free parameters and values of
$\chi^2/{\rm dof}$ are listed in Table \ref{tab1}. 

\begin{figure*}
\centering
\includegraphics[width=0.52\textwidth]{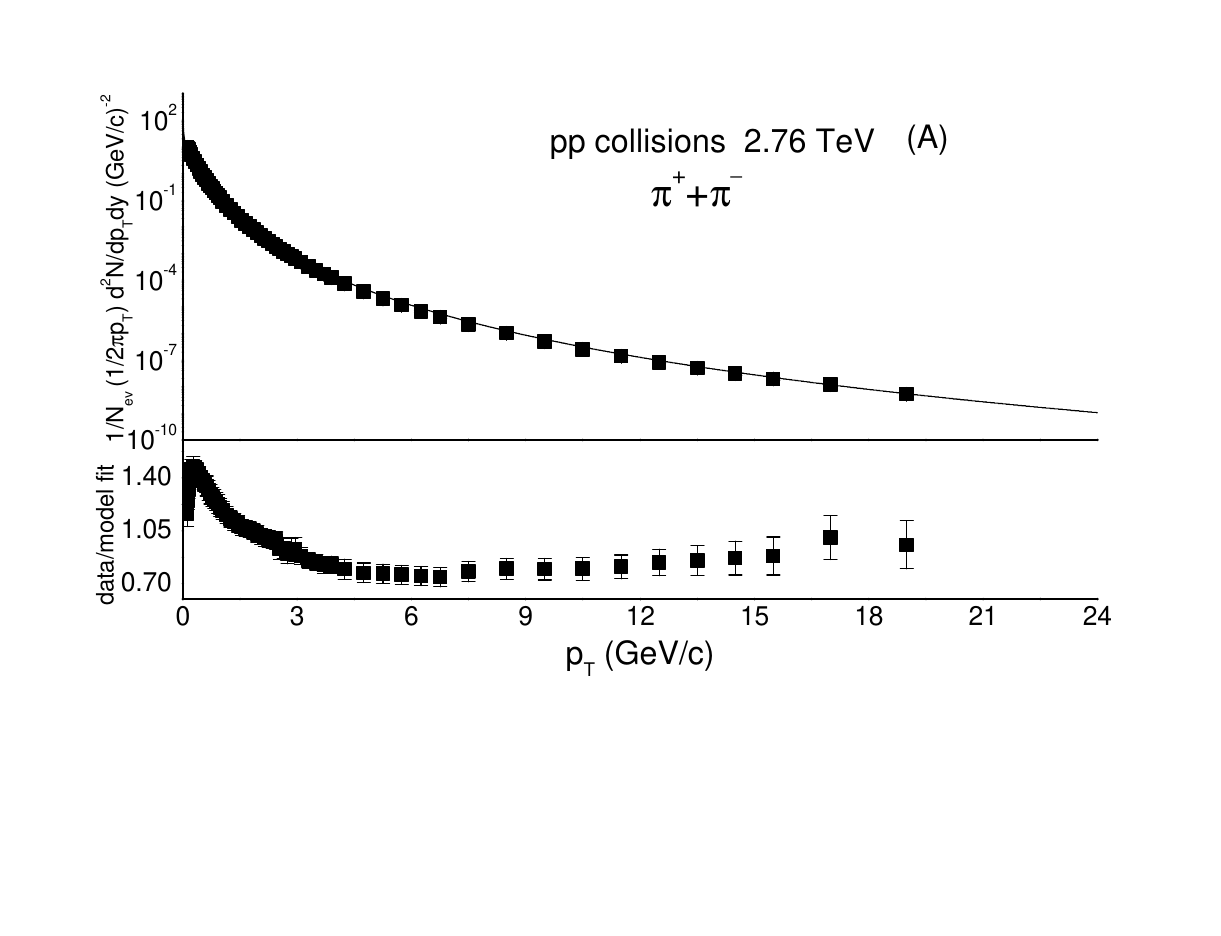}\hspace{-1.0cm}
\includegraphics[width=0.52\textwidth]{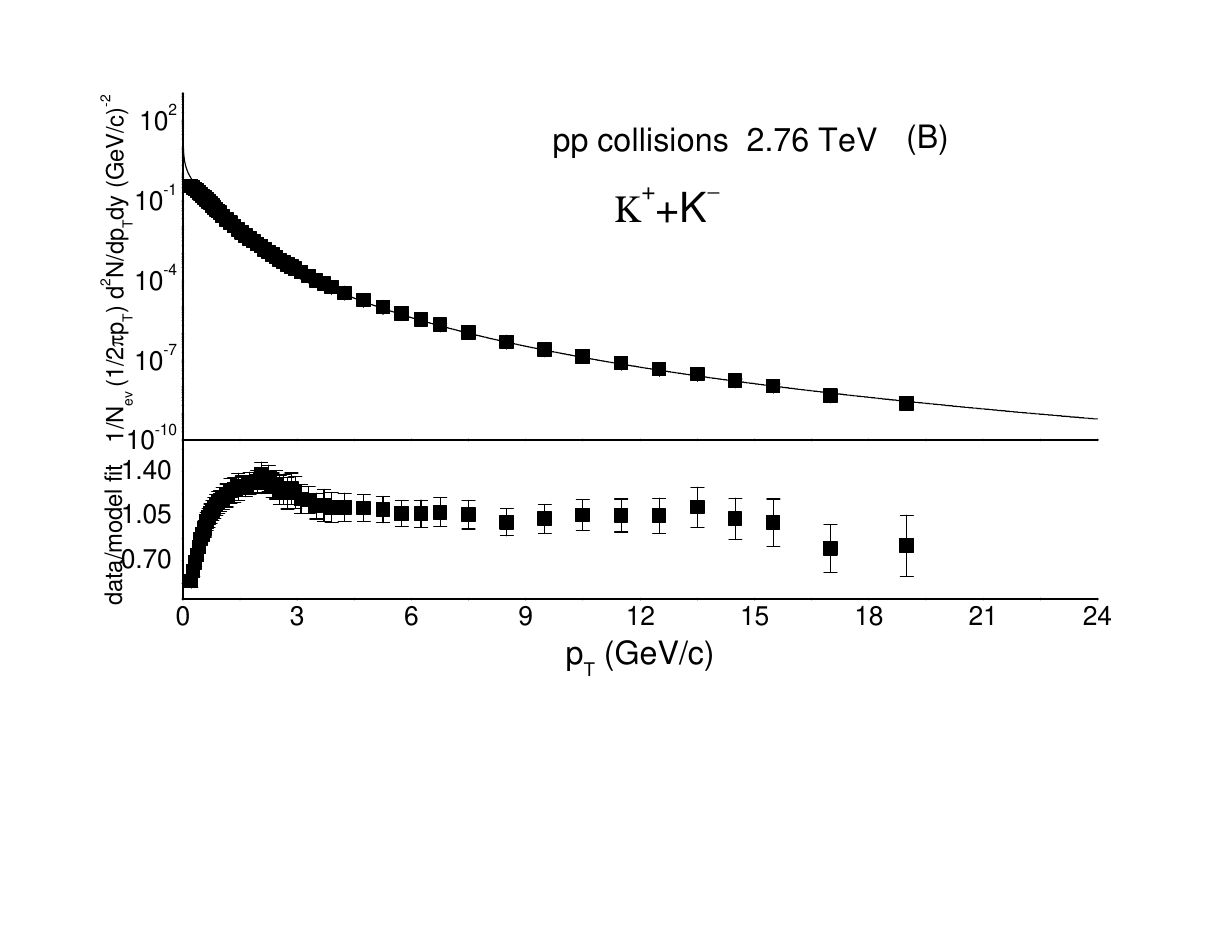}\hspace{-1.0cm}
\vspace*{-5mm} \\
\includegraphics[width=0.52\textwidth]{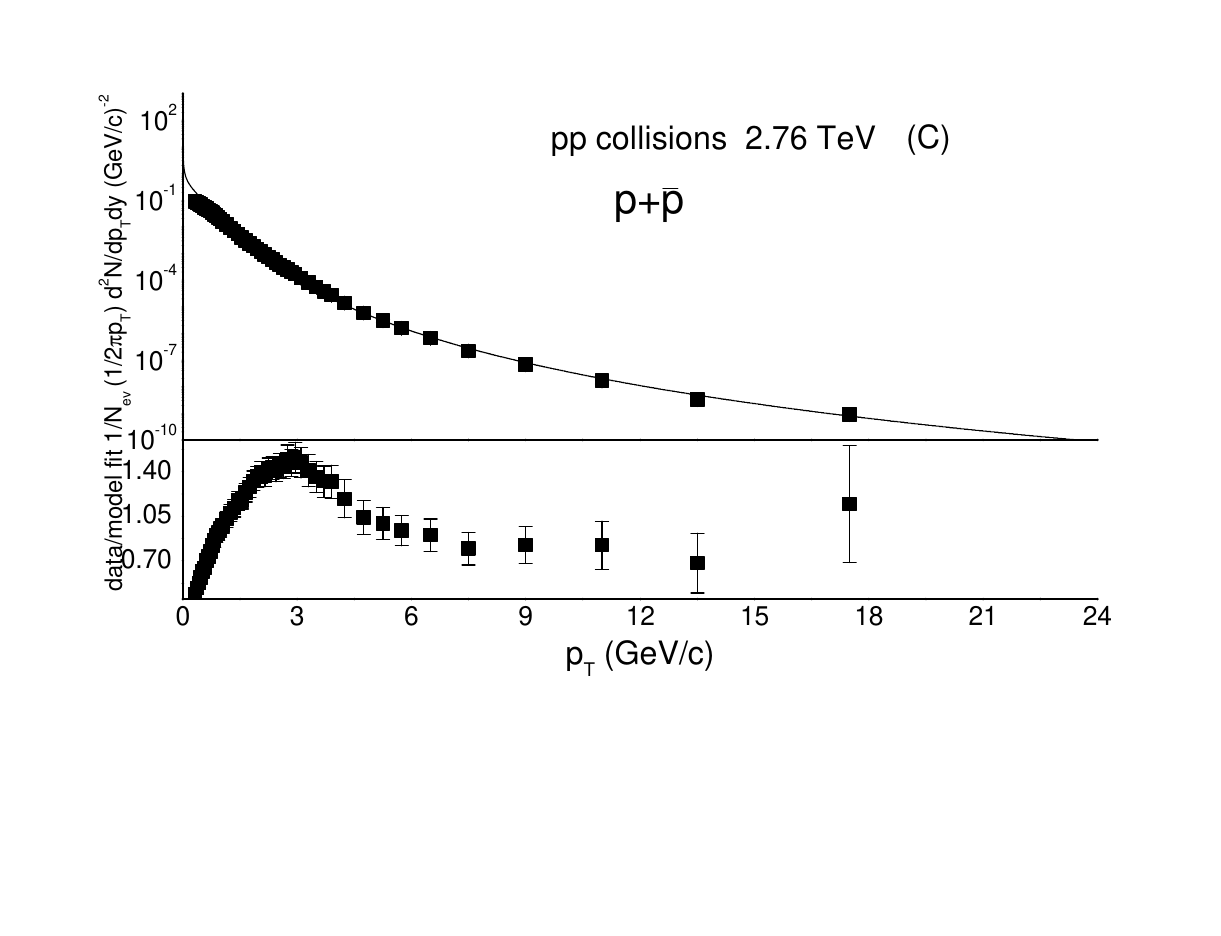}\hspace{-1.0cm}
\includegraphics[width=0.52\textwidth]{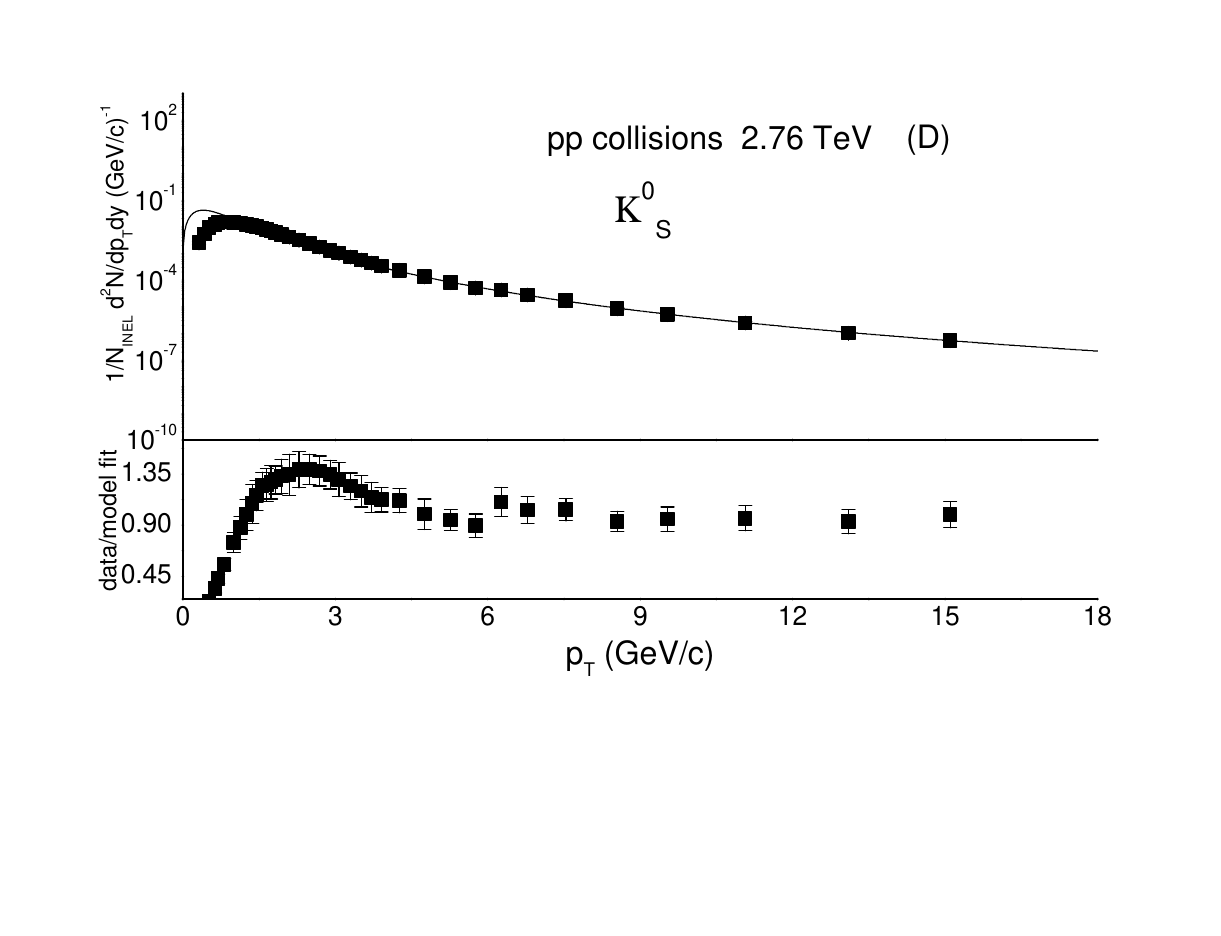}\hspace{-1.0cm}
\vspace*{-5mm} \\
\includegraphics[width=0.52\textwidth]{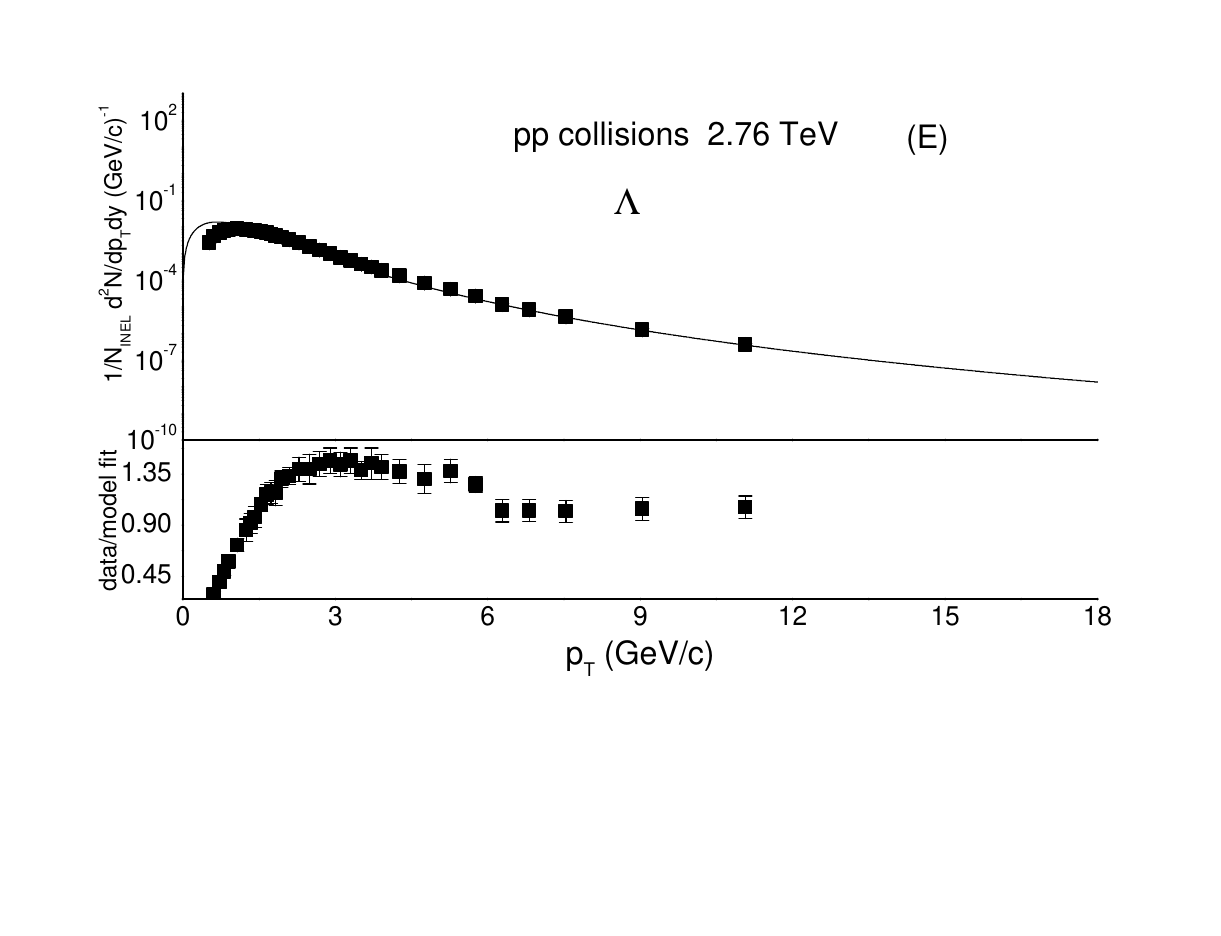}\hspace{-1.0cm}
\includegraphics[width=0.52\textwidth]{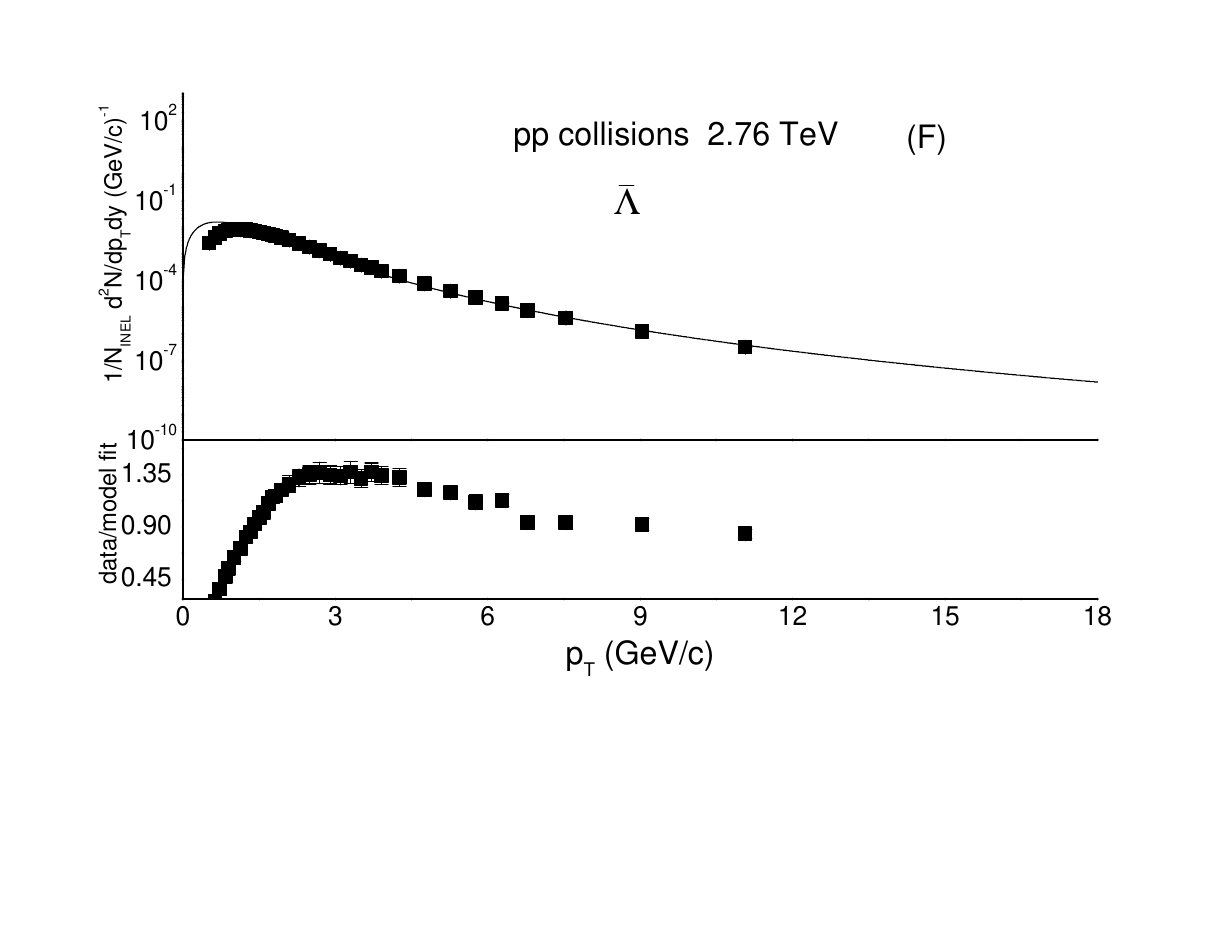}\hspace{-1.0cm}
\vspace*{-5mm} \\
\includegraphics[width=0.52\textwidth]{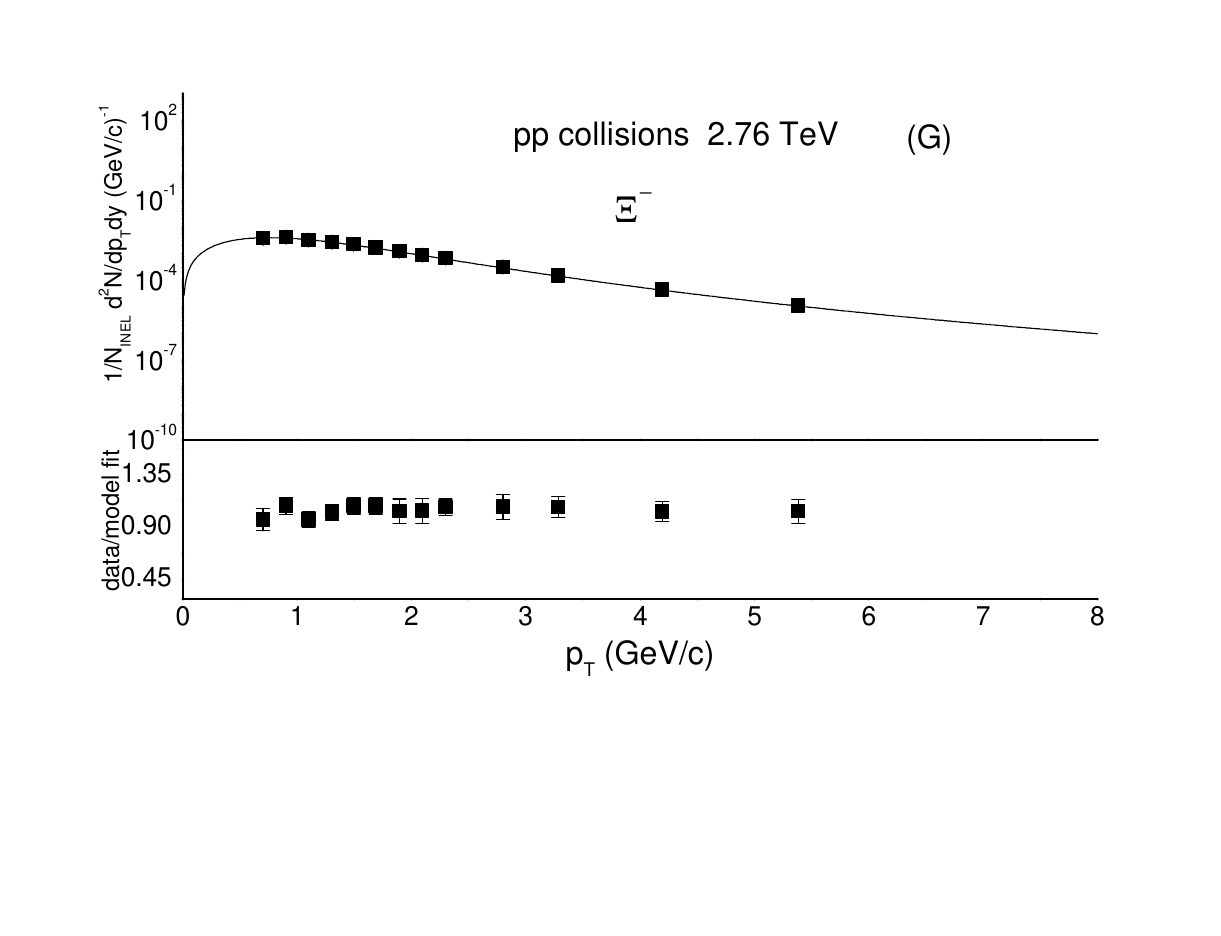}\hspace{-1.0cm}
\includegraphics[width=0.52\textwidth]{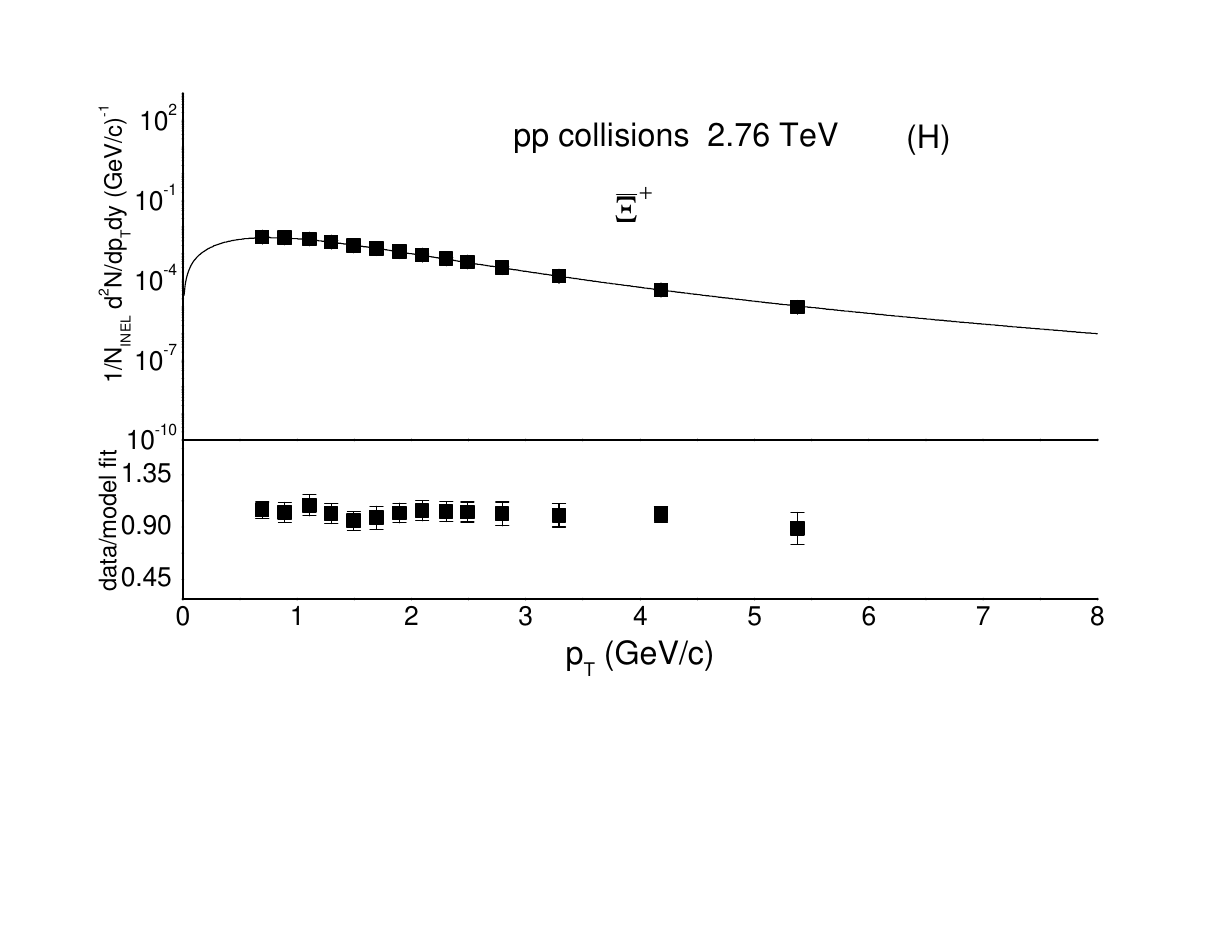}\hspace{-1.0cm}
\label{fig1-a}
\end{figure*}

\setcounter{figure}{0}

\begin{figure*}
\centering
  \includegraphics[width=0.52\textwidth]{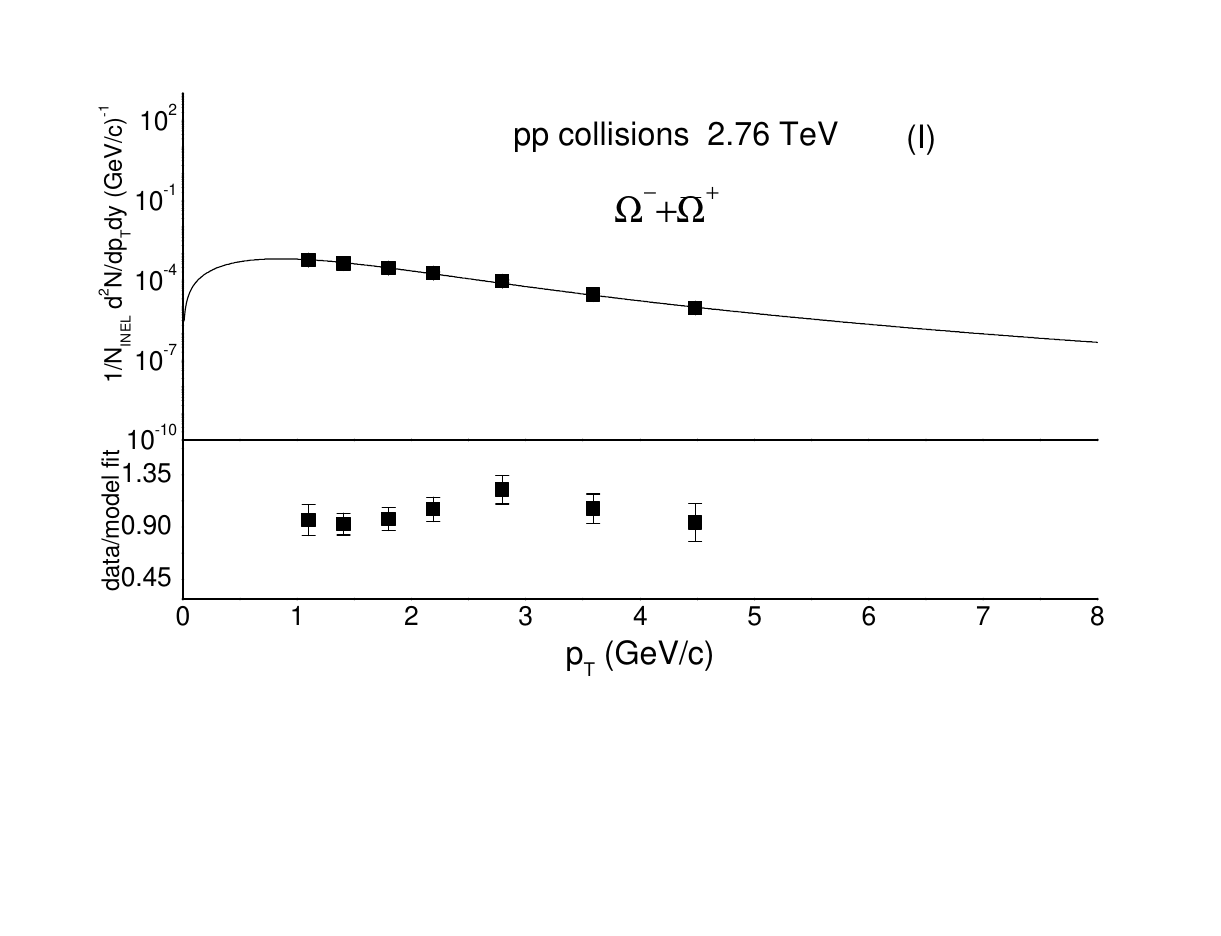}\hspace{-1.0cm}
  \includegraphics[width=0.52\textwidth]{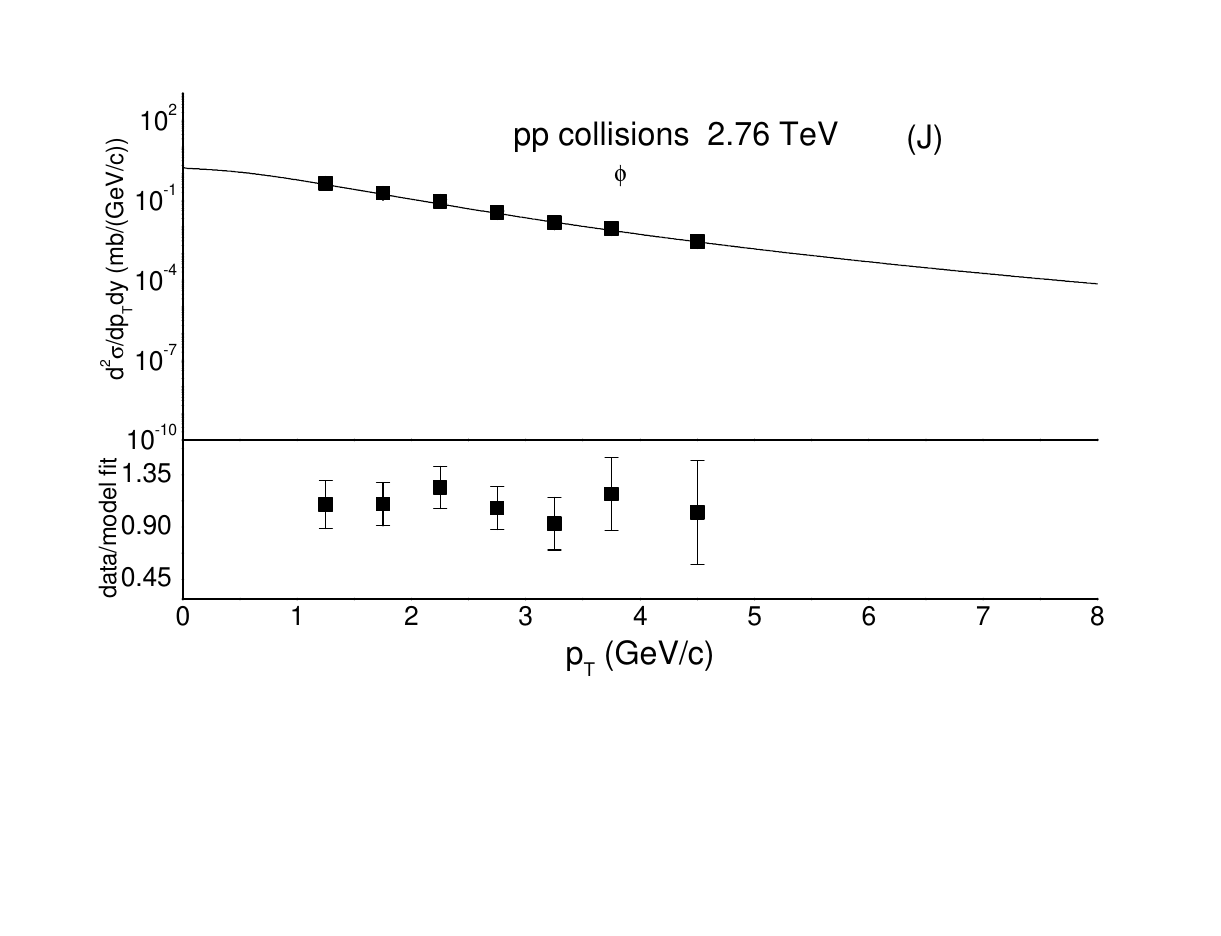}\hspace{-1.0cm}
\vspace*{-5mm} \\  
  \includegraphics[width=0.52\textwidth]{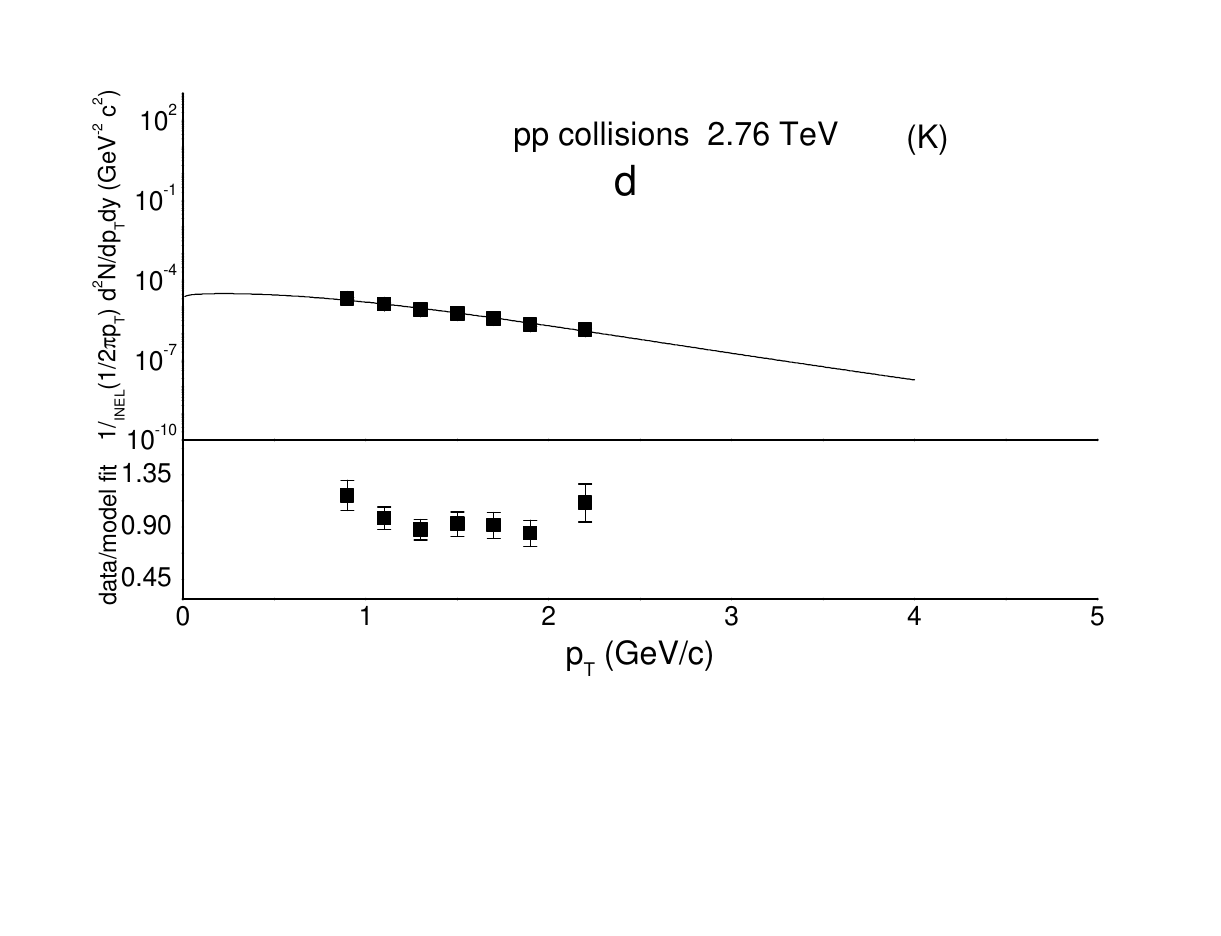}\hspace{-1.0cm}
  \includegraphics[width=0.52\textwidth]{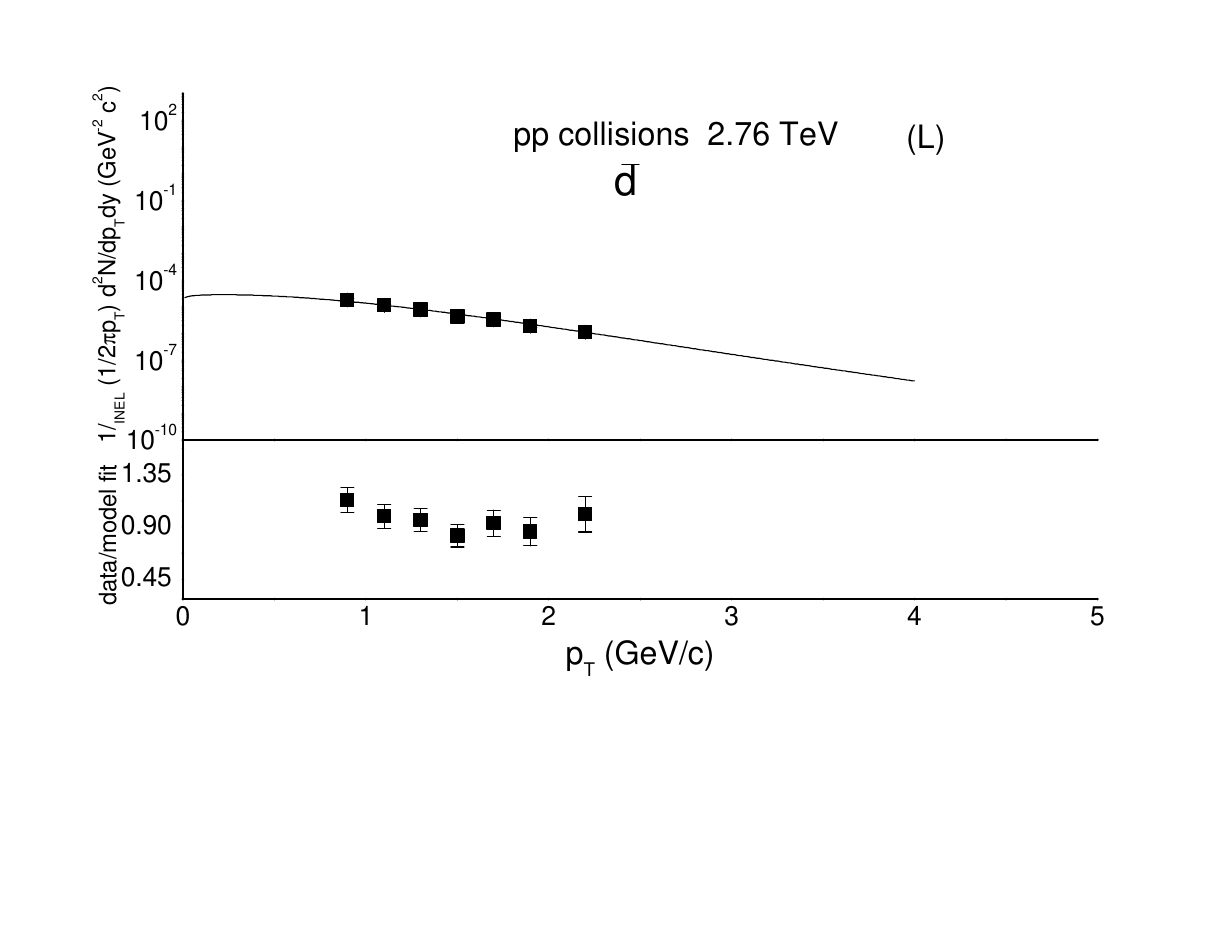}\hspace{-1.0cm}
  \vspace*{-10mm}
  \caption{The \(p_T\) spectra of a multitude of particles produced
    in $pp$ collisions at 2.76 TeV. The panels refer to different
    particle specie. The experimental data were taken from Refs.\
    \cite{ALICE:2014juv,Schuchmann:2015lay,Colella:2013iky,ALICE:2015cql}.
    The curves are fits to the function in Eq.\ (\ref{eq4}).
    The lower part of each panel shows the corresponding
    ratio data/fit.}
\label{fig1-b}
\end{figure*}

Figure \ref{fig2-b} is similar to Figure \ref{fig1-b},
but it shows the \(p_T\) spectra
of the particles at 7 TeV. It consists of panels (A) to (P), and
each panel shows one particle type. Panels (A) to (H) refer to the \(p_T\)
spectra of $\pi^\pm$, $K^\pm$, $p+\bar p$, $K^0_S$, $K^{*0}+\bar K^{*0}$,
$\Lambda + \bar \Lambda$, $\Xi^- + \bar \Xi^+$ and
$\Omega^- + \bar \Omega^+$, respectively, while $\phi$, $d$,
$\bar d$, triton $t$, anti-triton $\bar t$, \(^3\text{He}\)
and \(^3\overline{\mathrm{He}}\)
are shown in panel (I) to (P). Again, the experimental
data of the ALICE Collaboration are are taken from Refs.\
\cite{ALICE:2020jsh,ALICE:2012wos,ALICE:2017xrp}. 

\begin{figure*}
\centering
\includegraphics[width=0.52\textwidth]{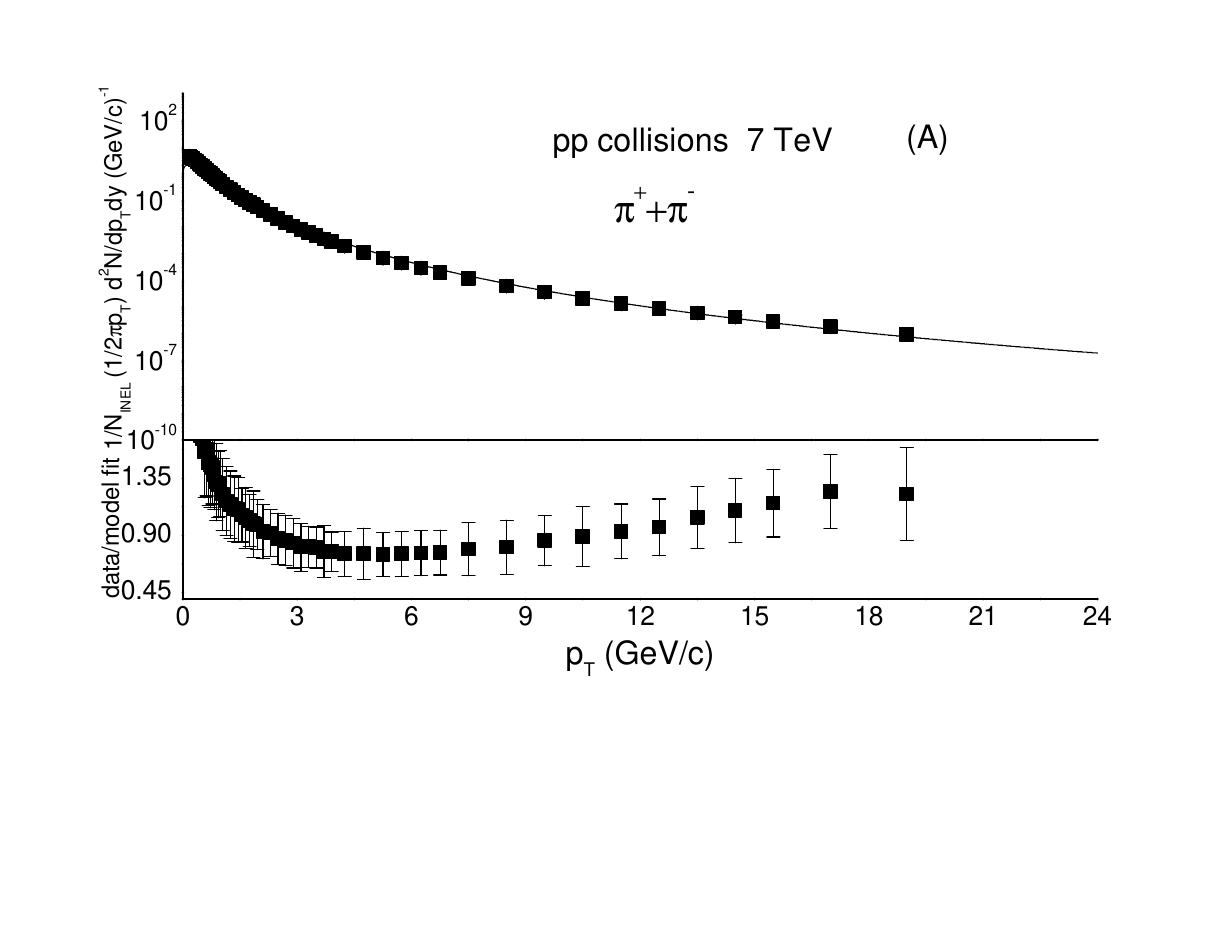}\hspace{-1.0cm}
\includegraphics[width=0.52\textwidth]{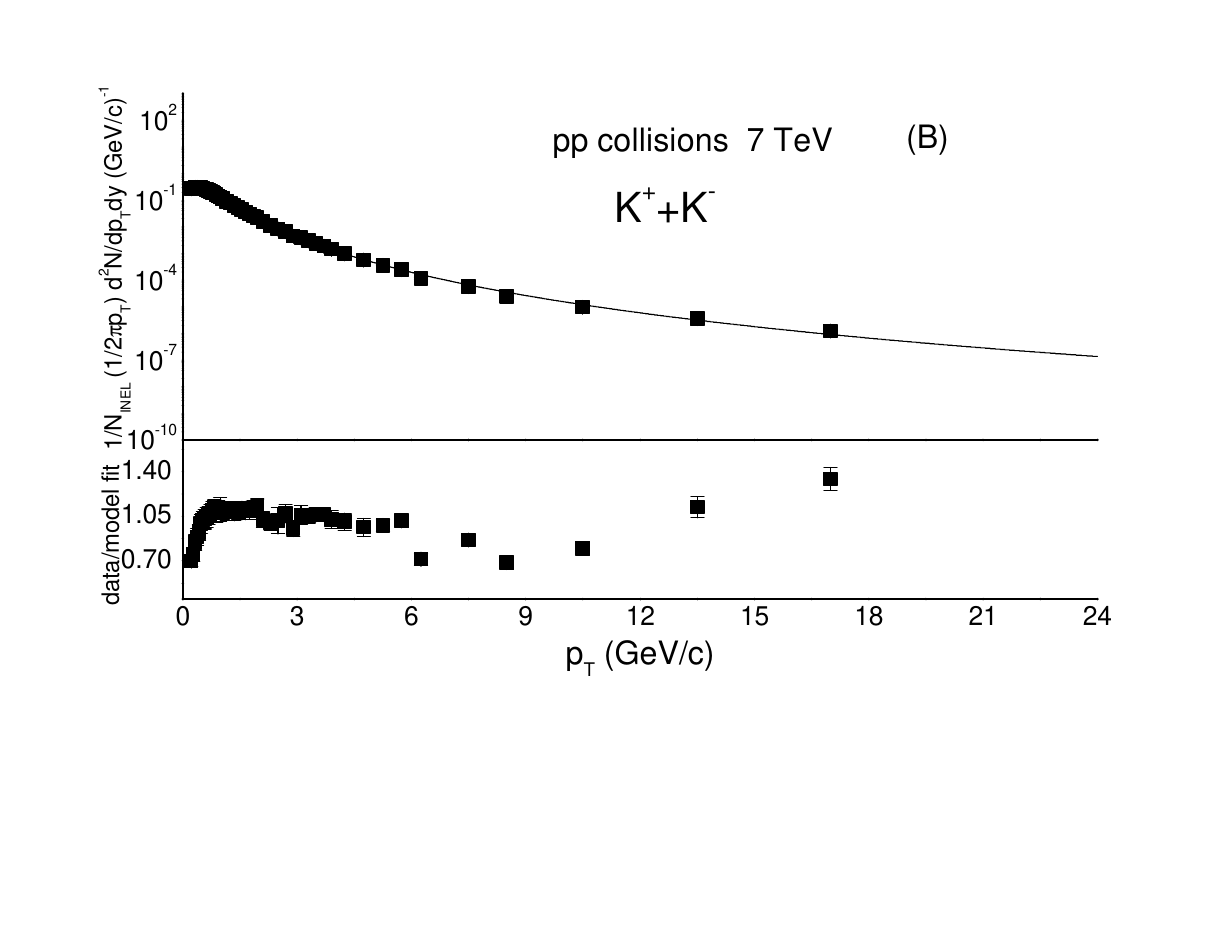}\hspace{-1.0cm}
\vspace*{-5mm} \\
\includegraphics[width=0.52\textwidth]{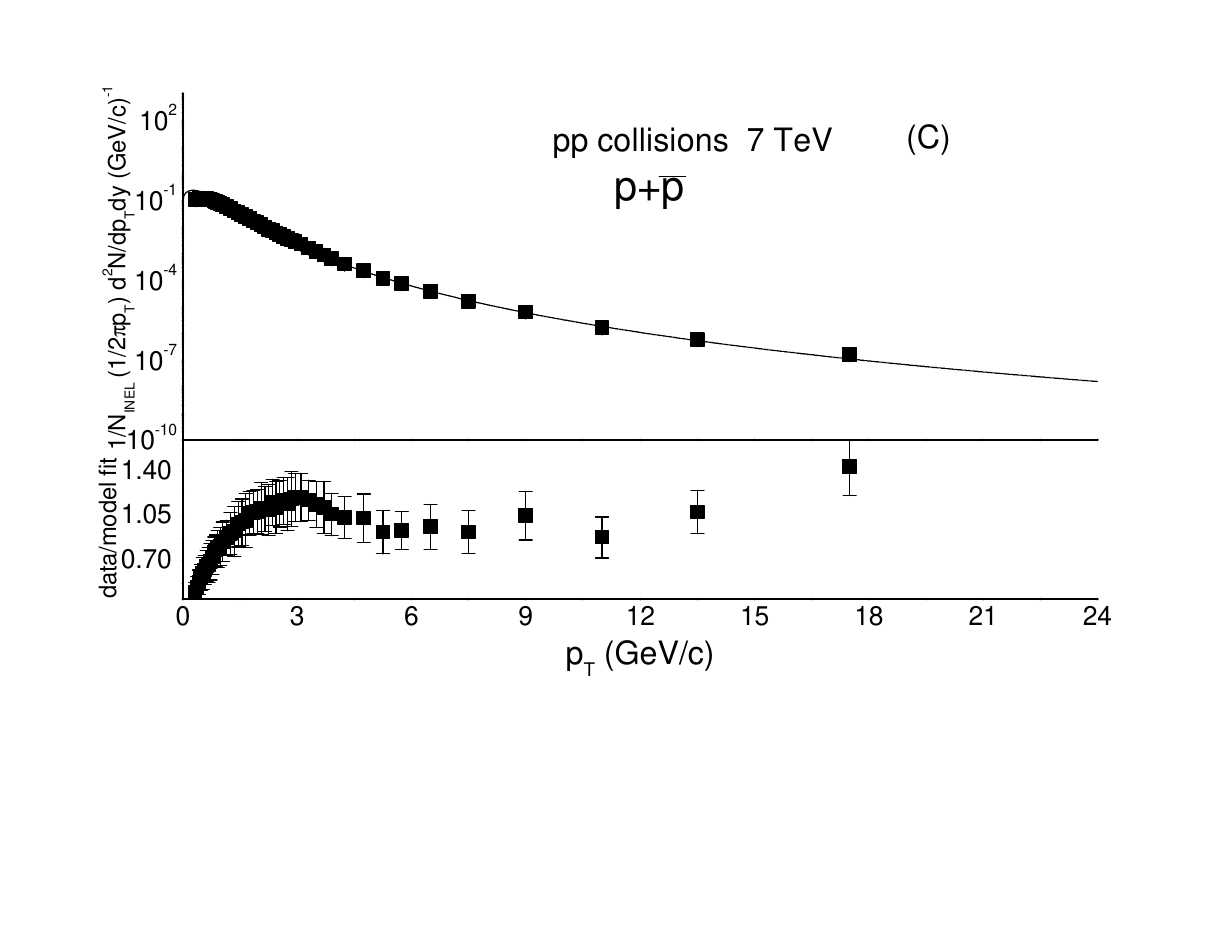}\hspace{-1.0cm}
\includegraphics[width=0.52\textwidth]{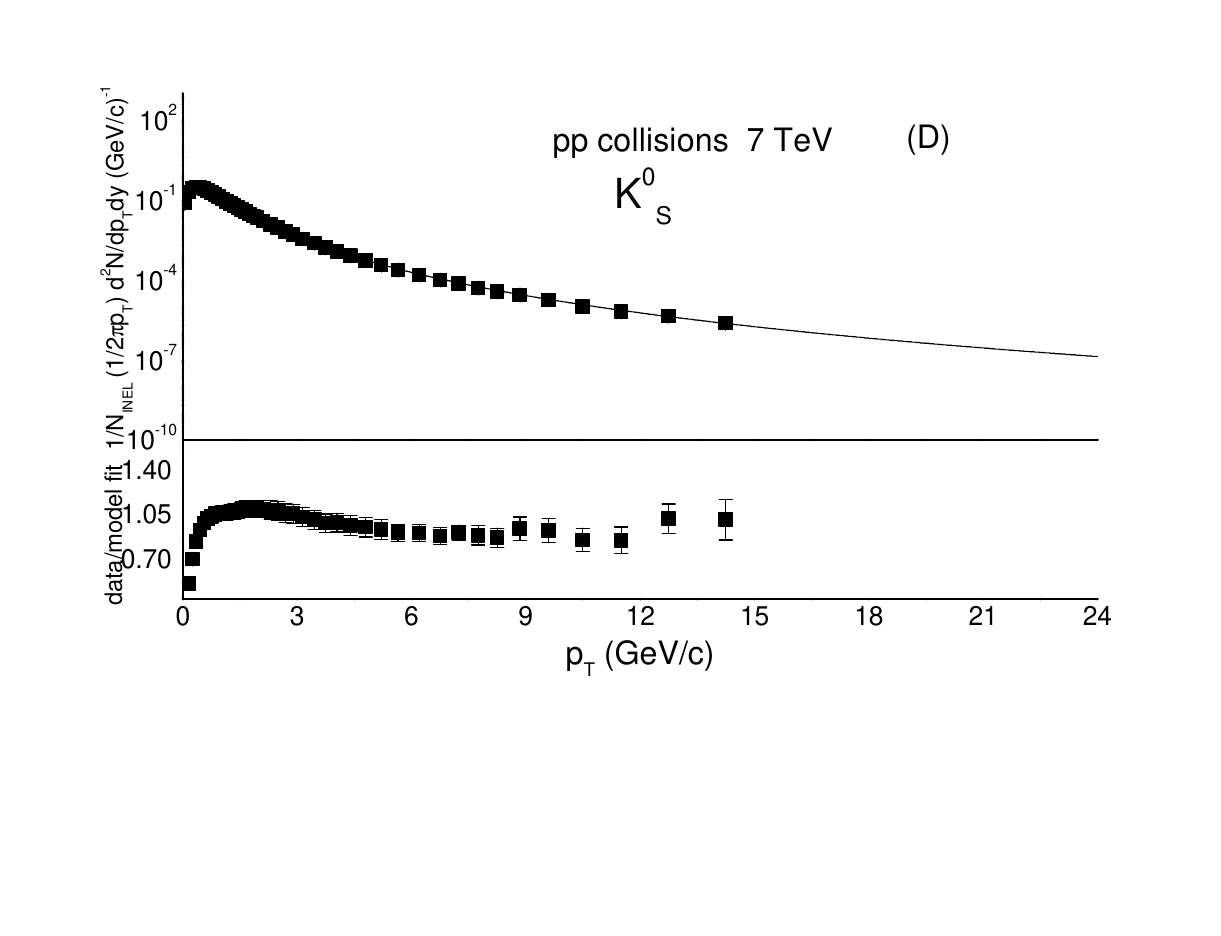}\hspace{-1.0cm}
\vspace*{-5mm} \\
\includegraphics[width=0.52\textwidth]{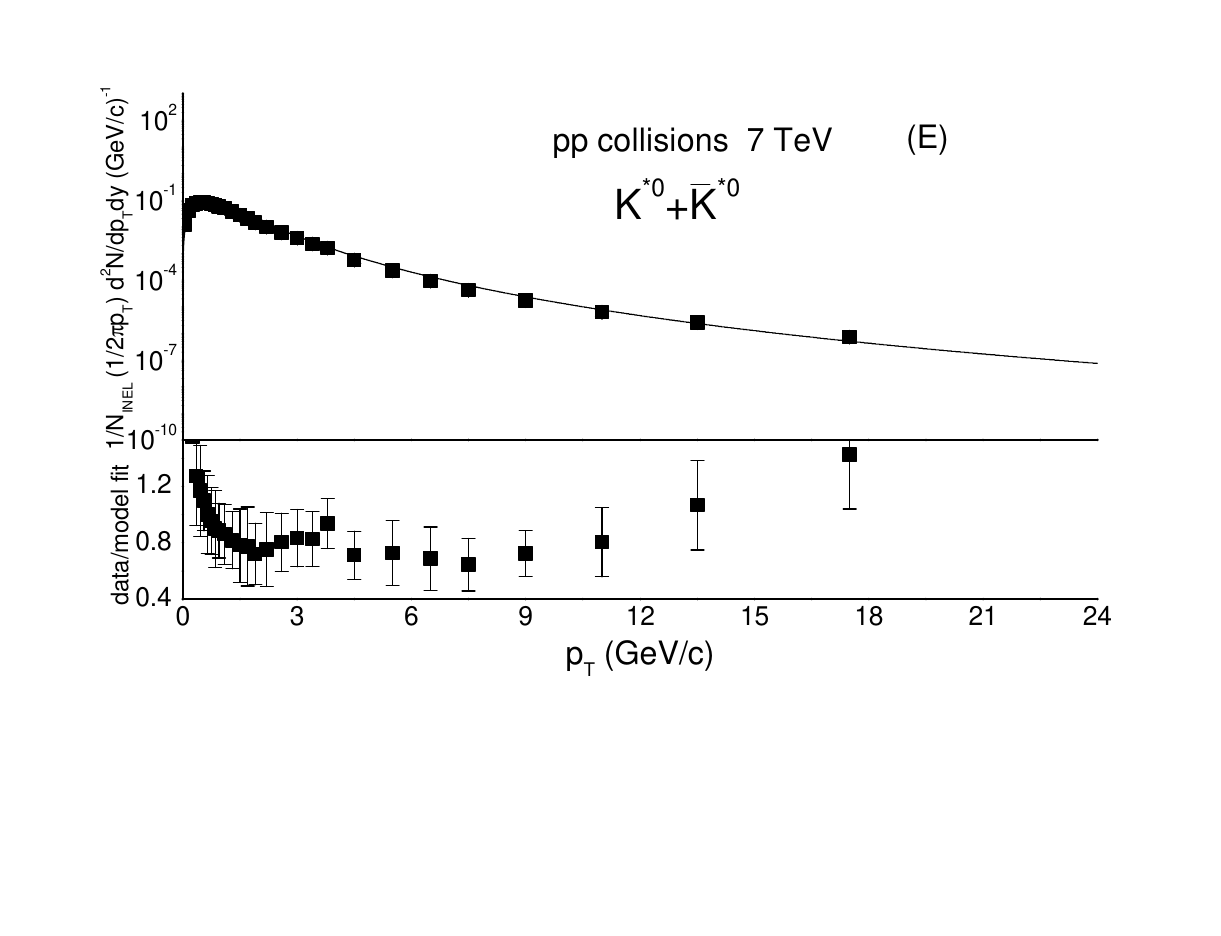}\hspace{-1.0cm}
\includegraphics[width=0.52\textwidth]{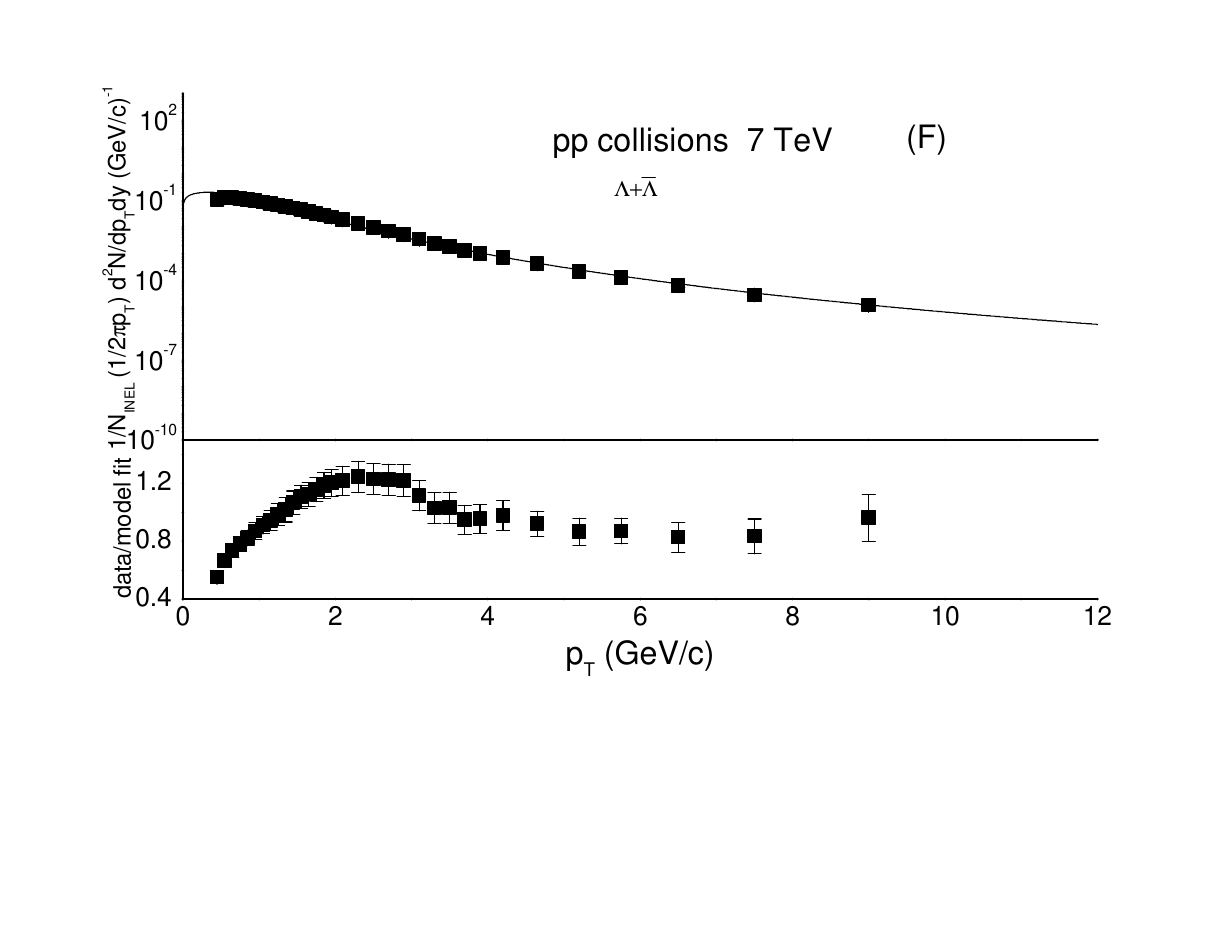}\hspace{-1.0cm}
\vspace*{-5mm} \\
\includegraphics[width=0.52\textwidth]{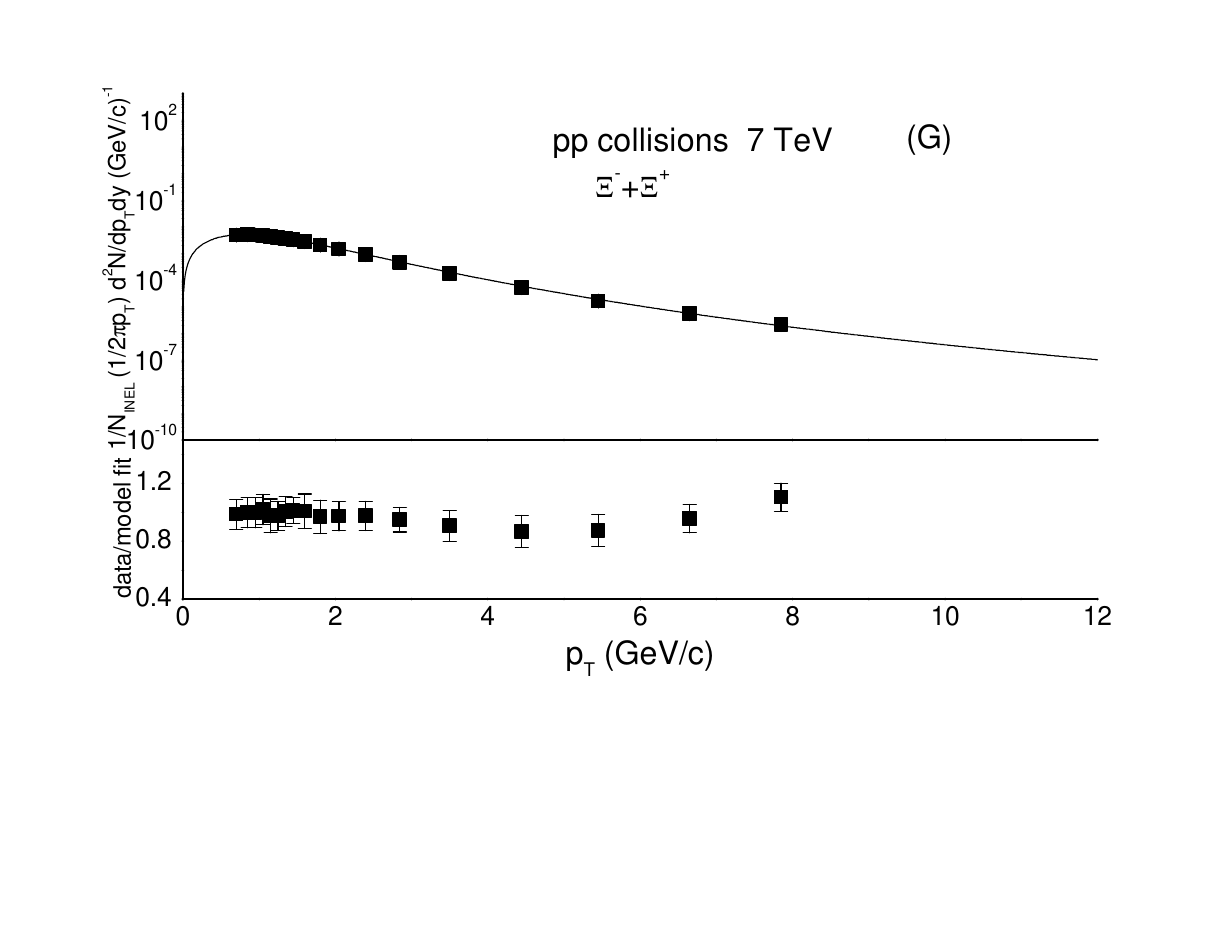}\hspace{-1.0cm}
\includegraphics[width=0.52\textwidth]{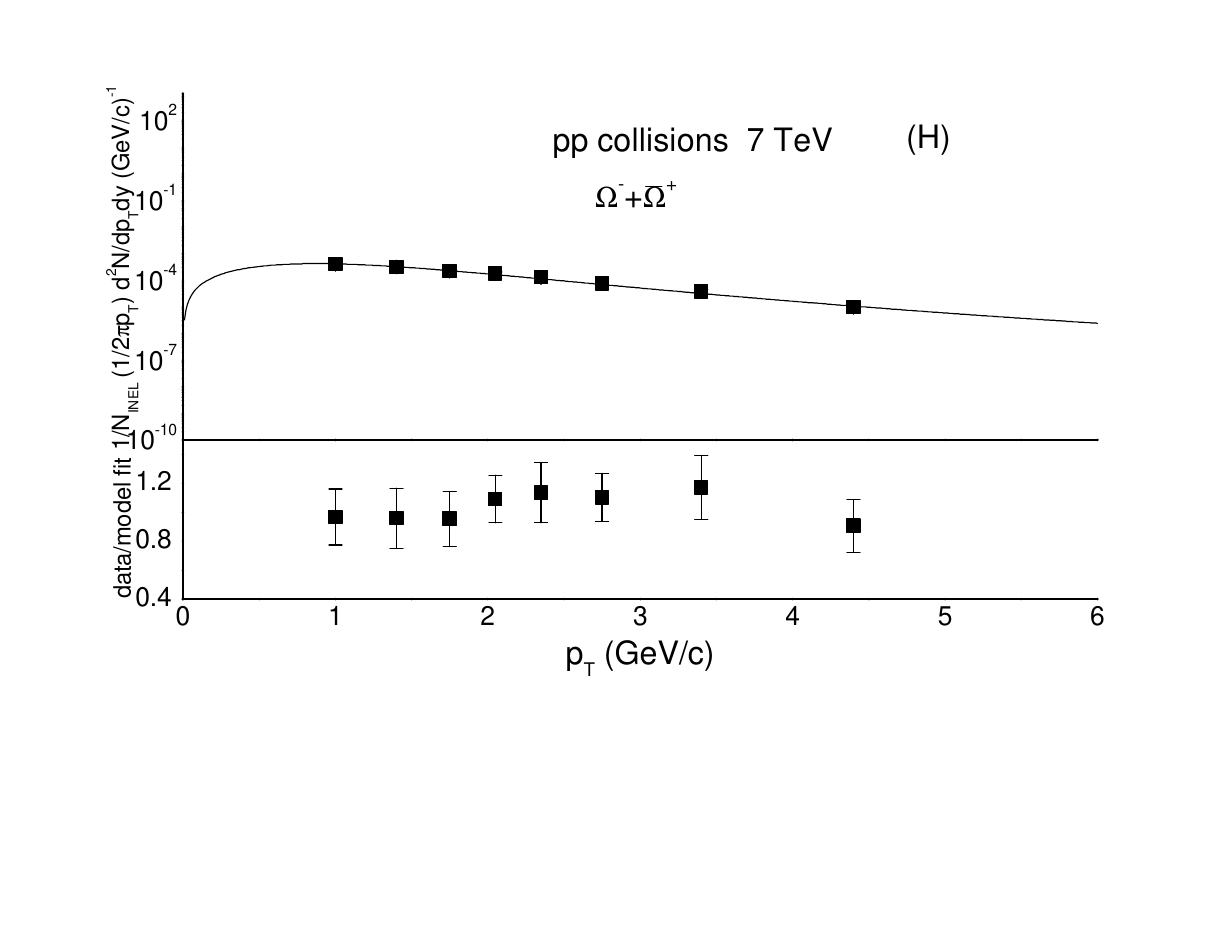}\hspace{-1.0cm}
\label{fig2-a}
\end{figure*}

\setcounter{figure}{1}
\begin{figure*}
\centering
  \includegraphics[width=0.45\textwidth]{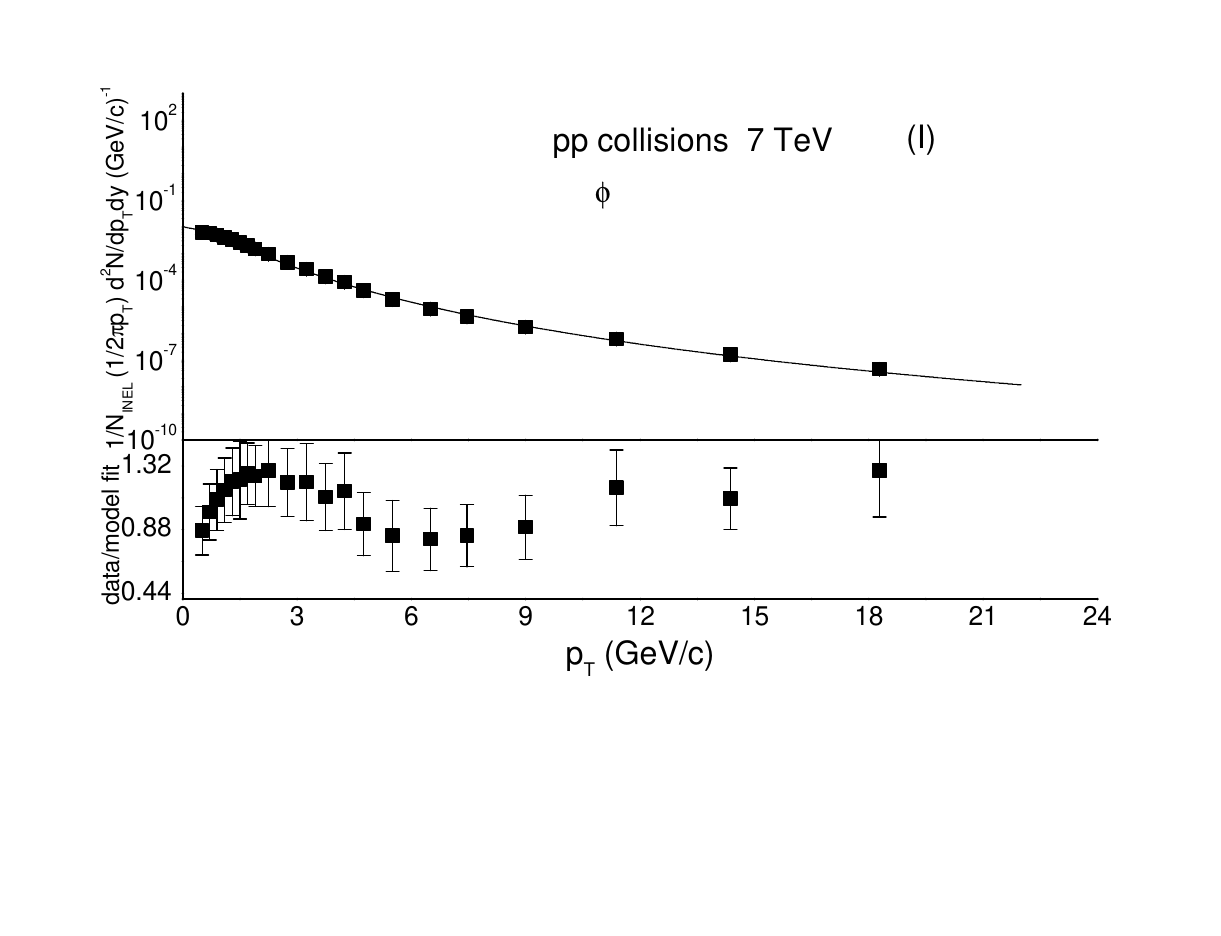}\hspace{-0.6cm}
  \includegraphics[width=0.45\textwidth]{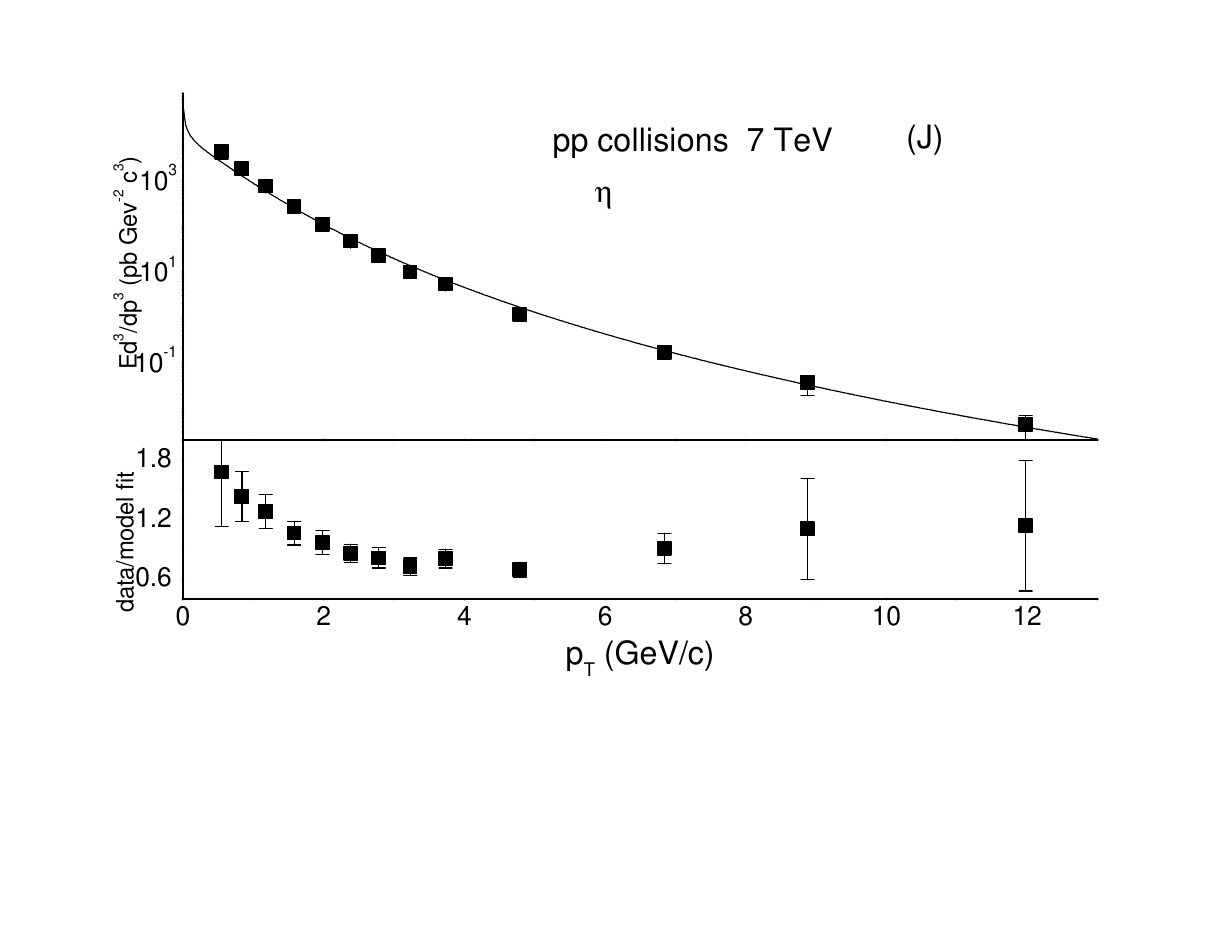}\hspace{-1.0cm}
\vspace*{-8mm} \\
  \includegraphics[width=0.45\textwidth]{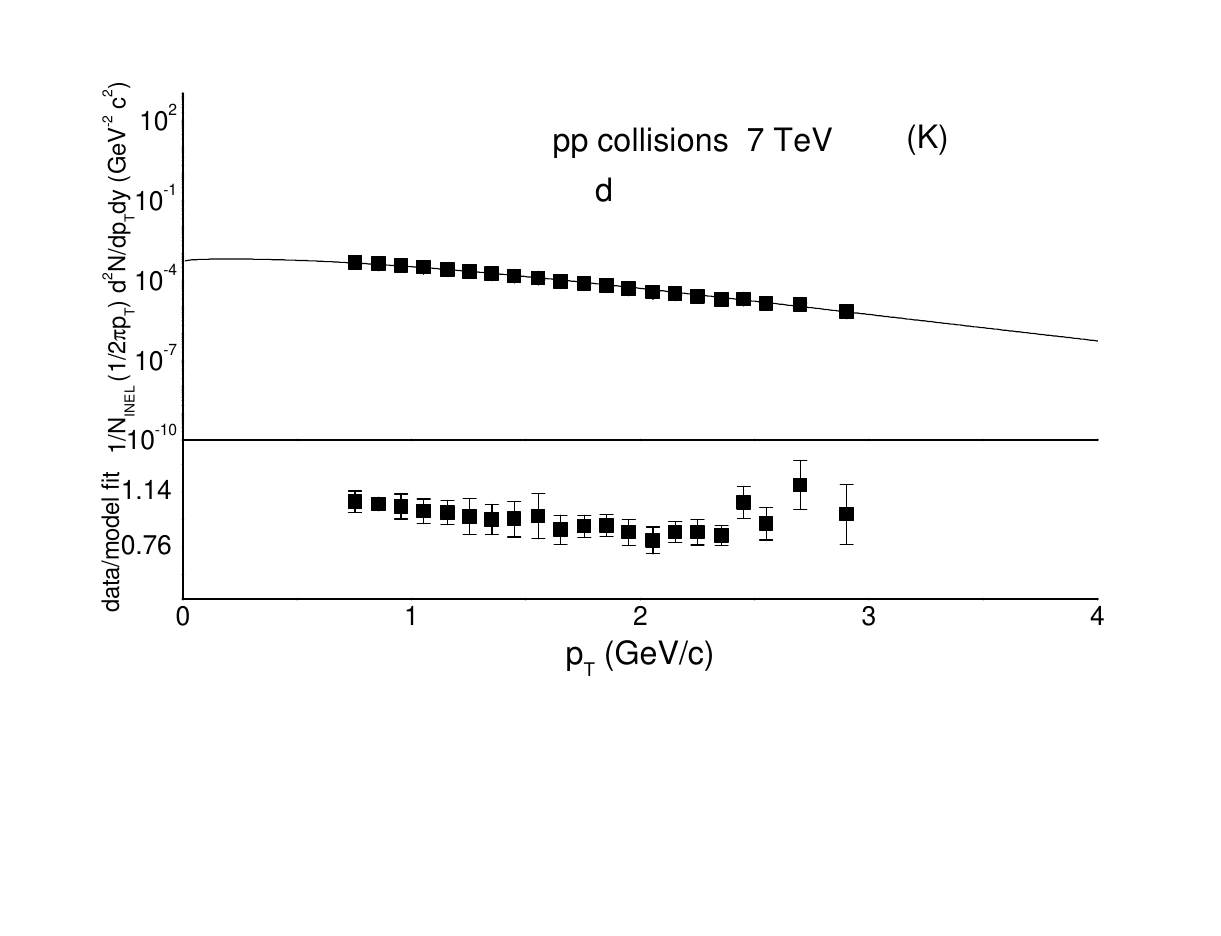}\hspace{-0.6cm}
  \includegraphics[width=0.45\textwidth]{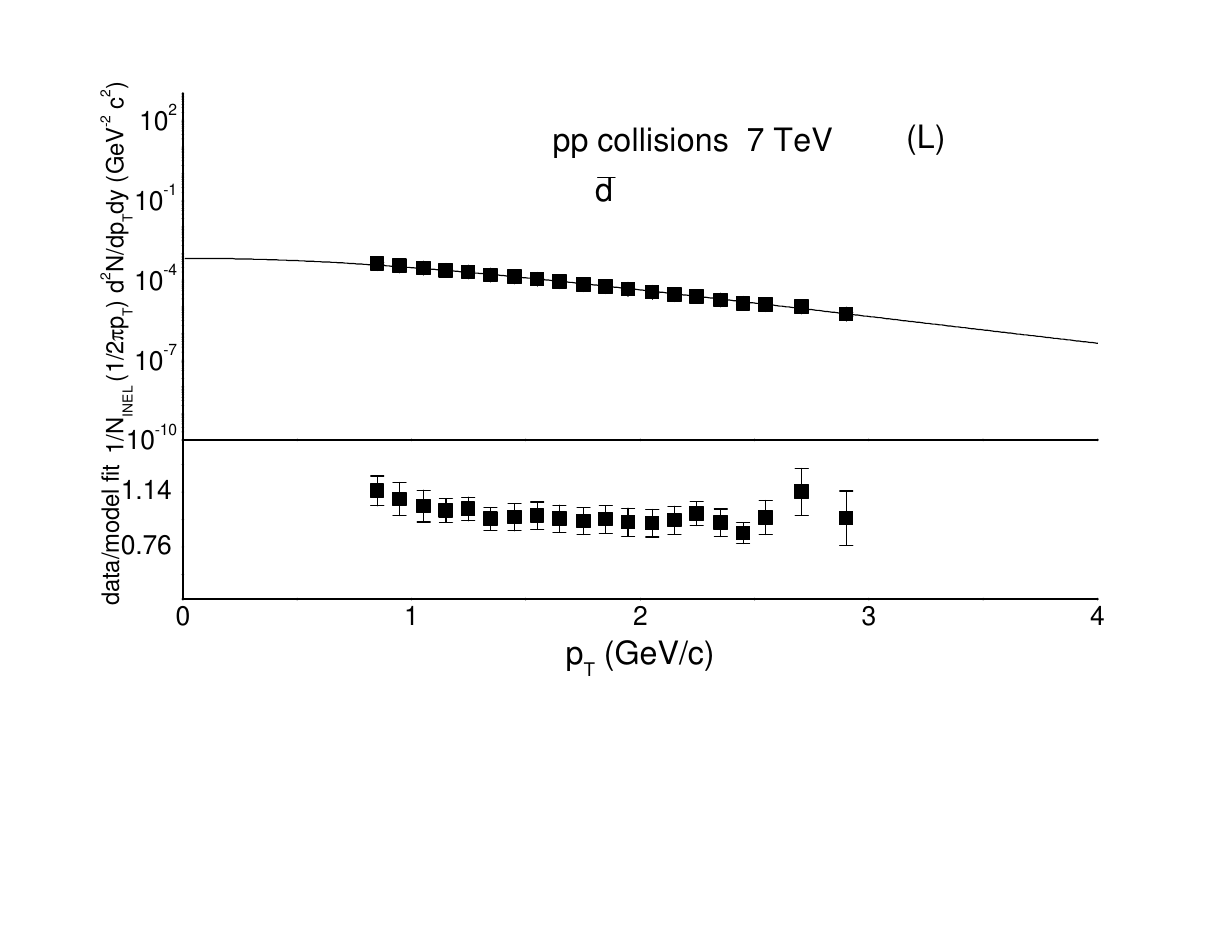}\hspace{-1.0cm}
\vspace*{-8mm} \\
  \includegraphics[width=0.45\textwidth]{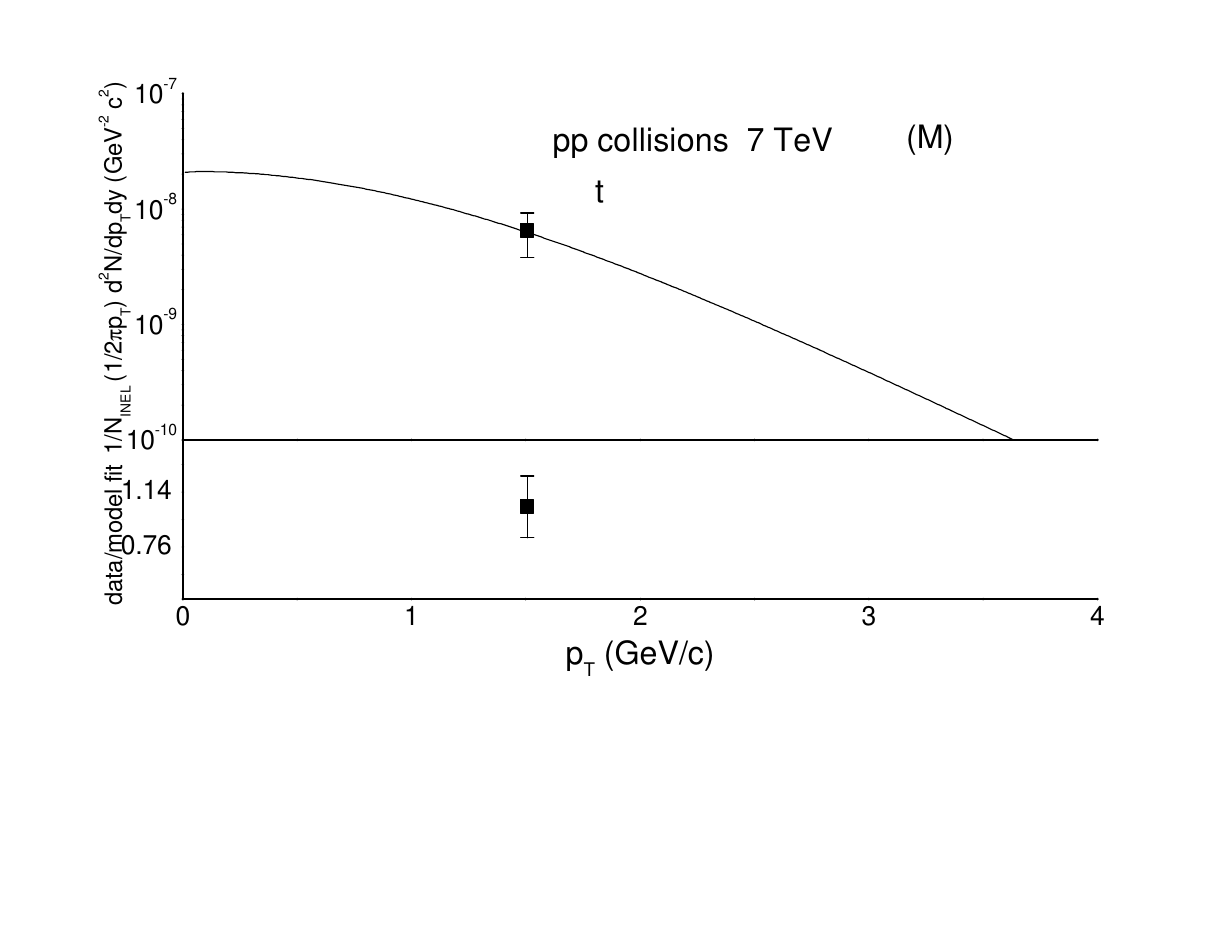}\hspace{-0.6cm}
  \includegraphics[width=0.45\textwidth]{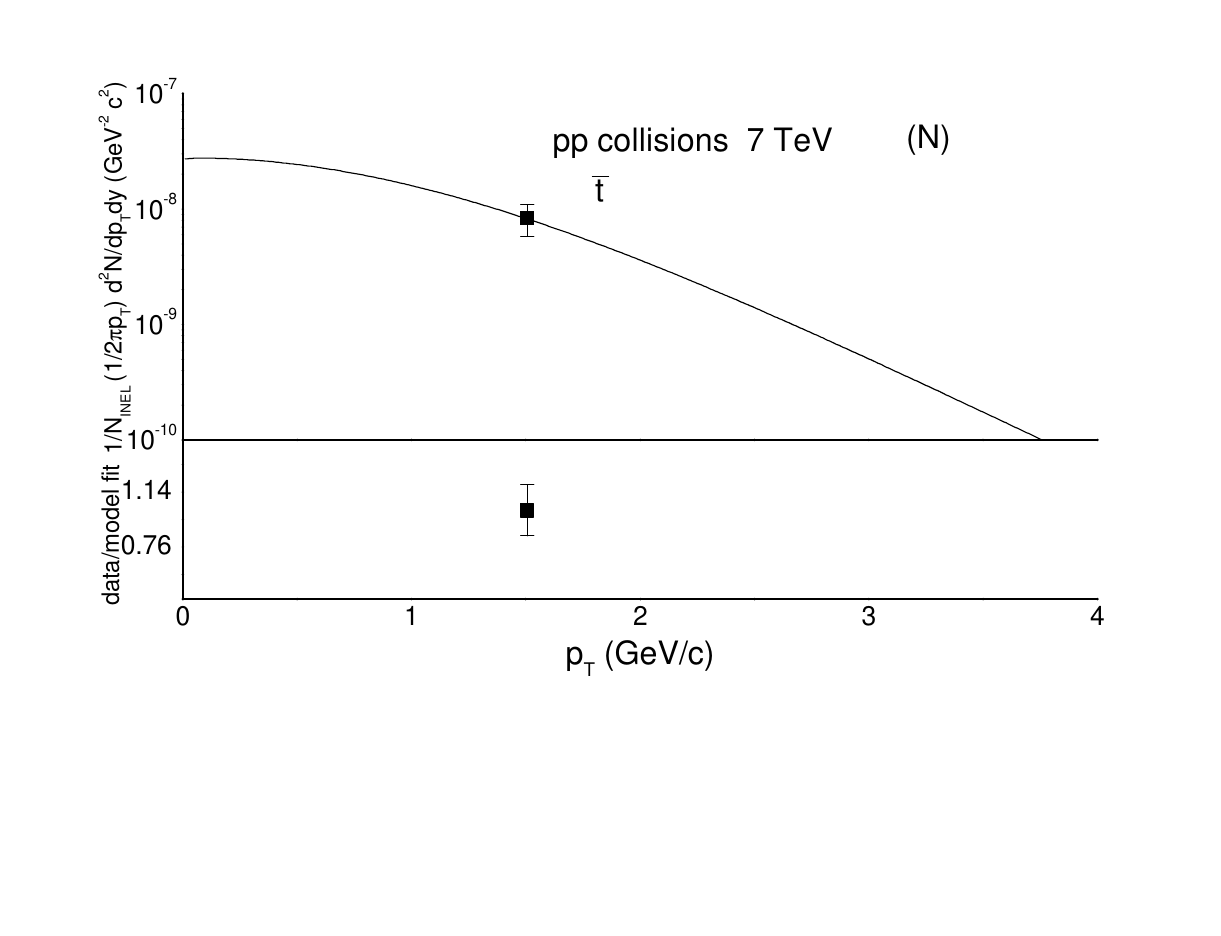}\hspace{-1.0cm}
\vspace*{-8mm}  \\
  \includegraphics[width=0.45\textwidth]{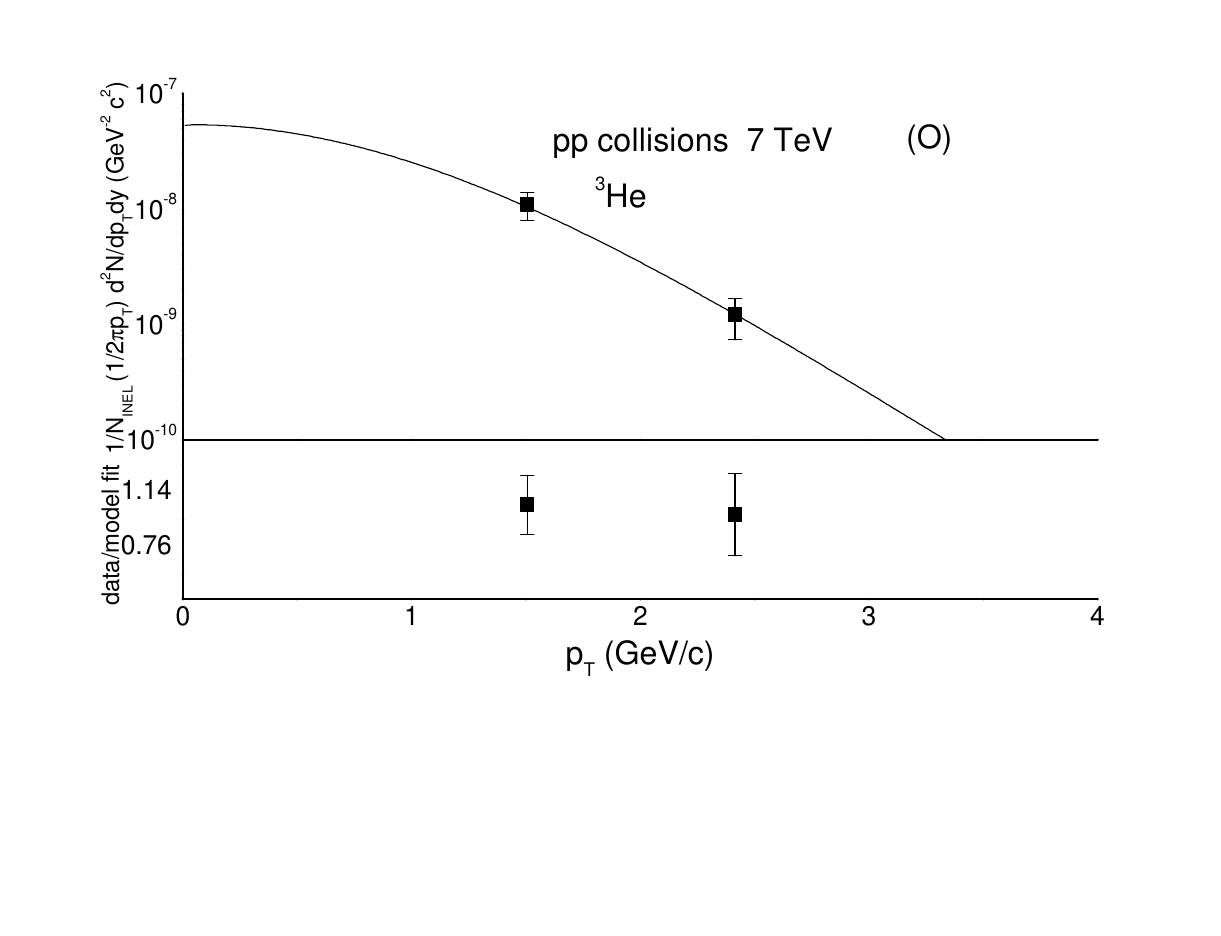}\hspace{-0.6cm}
  \includegraphics[width=0.45\textwidth]{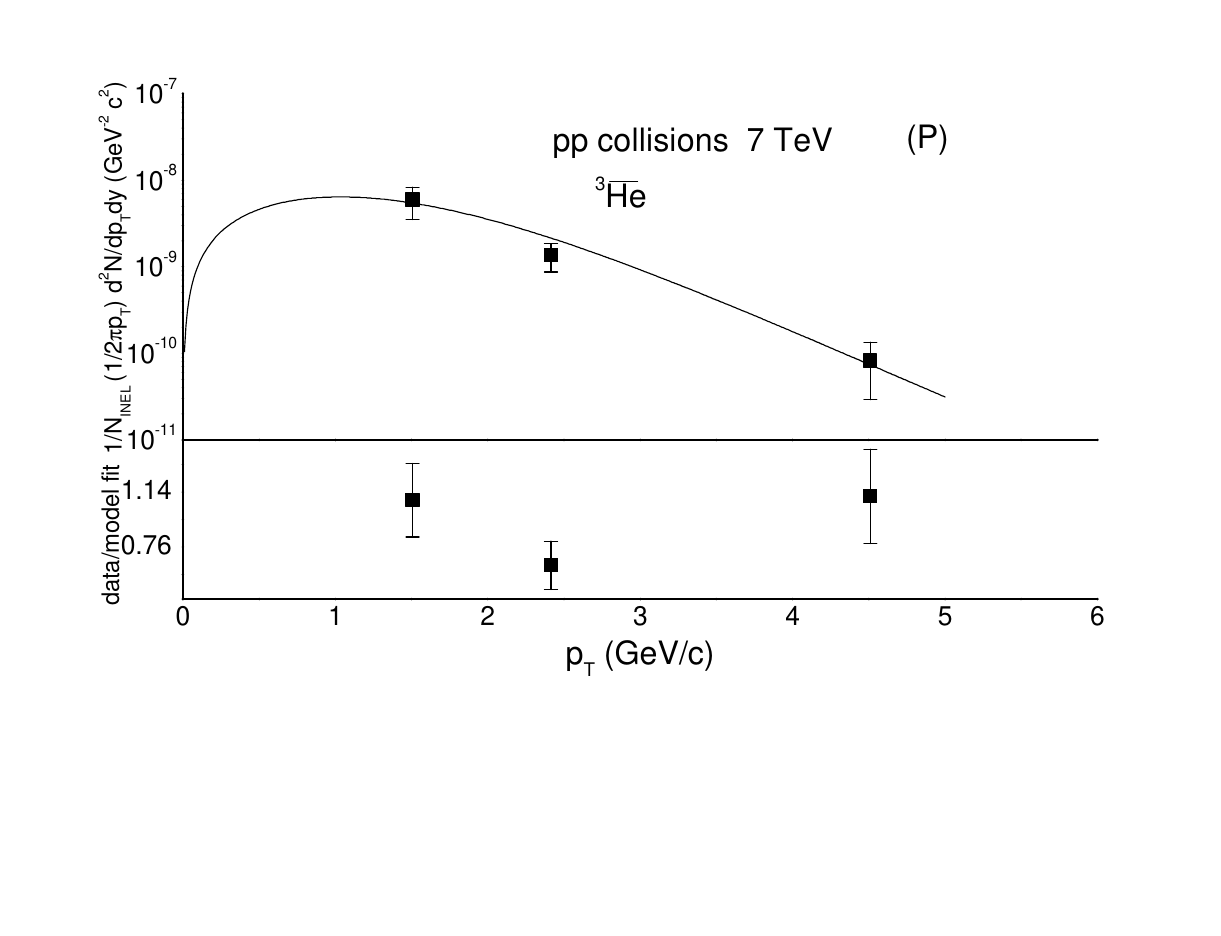}\hspace{-1.0cm}
  \vspace*{-10mm}
  \caption{This figure is similar to Figure \ref{fig1-b}, but it shows
    the \(p_T\) spectra of $\pi^\pm$, $K^\pm$, $p+\bar p$, $K^0_S$,
    $K^{*0}+\bar K^{*0}$, $\Lambda+\bar \Lambda$, $\Xi^-+\bar \Xi^+$,
    $\Omega^-+\bar \Omega^+$, $\phi$, $\eta$, $d$, $\bar d$, $t$,
    $\bar t$, \(^3\text{He}\) and \(^3\overline{\mathrm{He}}\) at
    $\sqrt{s} = 7~{\rm TeV}$. Each particle
    specie is presented separately in one of the panels
    (A) to (P). The experimental data of ALICE
    collaboration are taken from Refs.\
    \cite{ALICE:2020jsh,ALICE:2012wos,ALICE:2017xrp}.}  
\label{fig2-b}
\end{figure*}

\begin{figure*}
\vspace*{-5mm}
\centering
\hspace*{-10mm}
\includegraphics[width=0.55\textwidth]{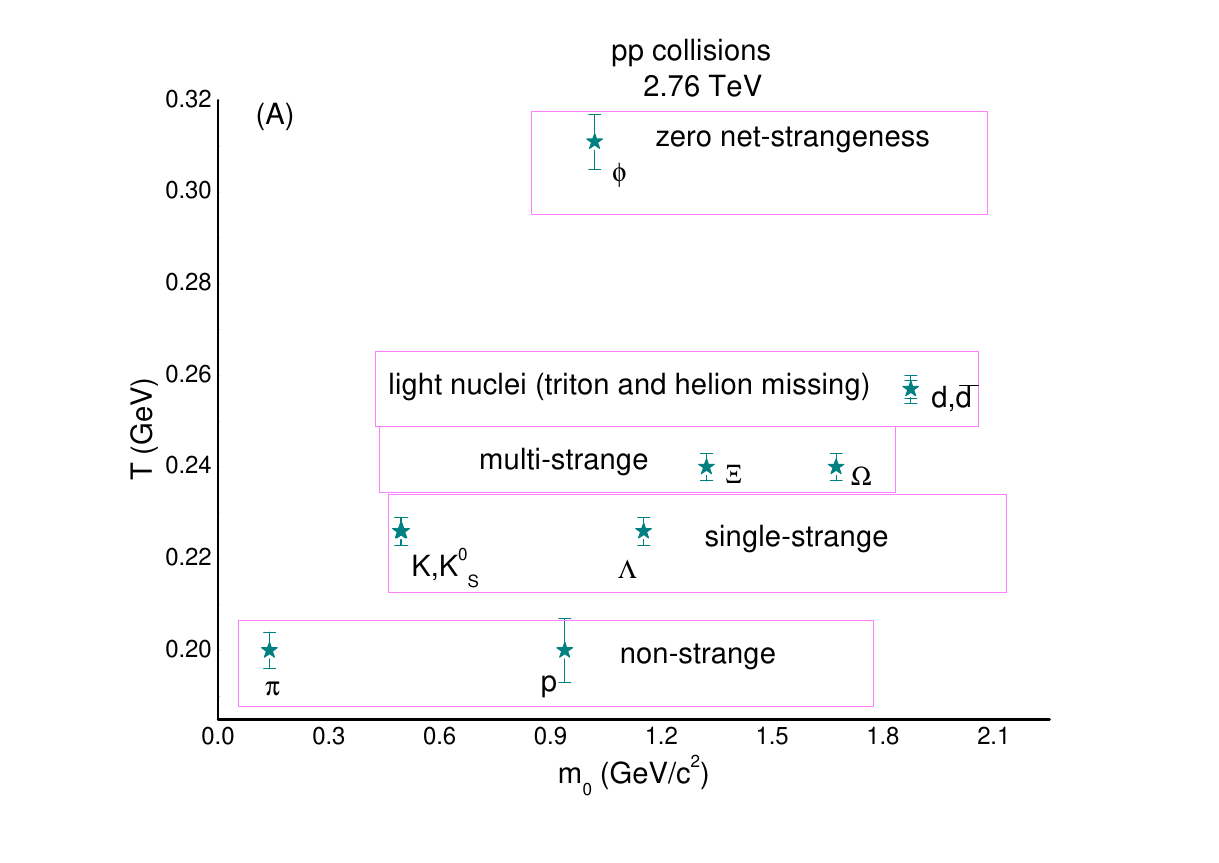}\hspace{-1.5cm}
\includegraphics[width=0.55\textwidth]{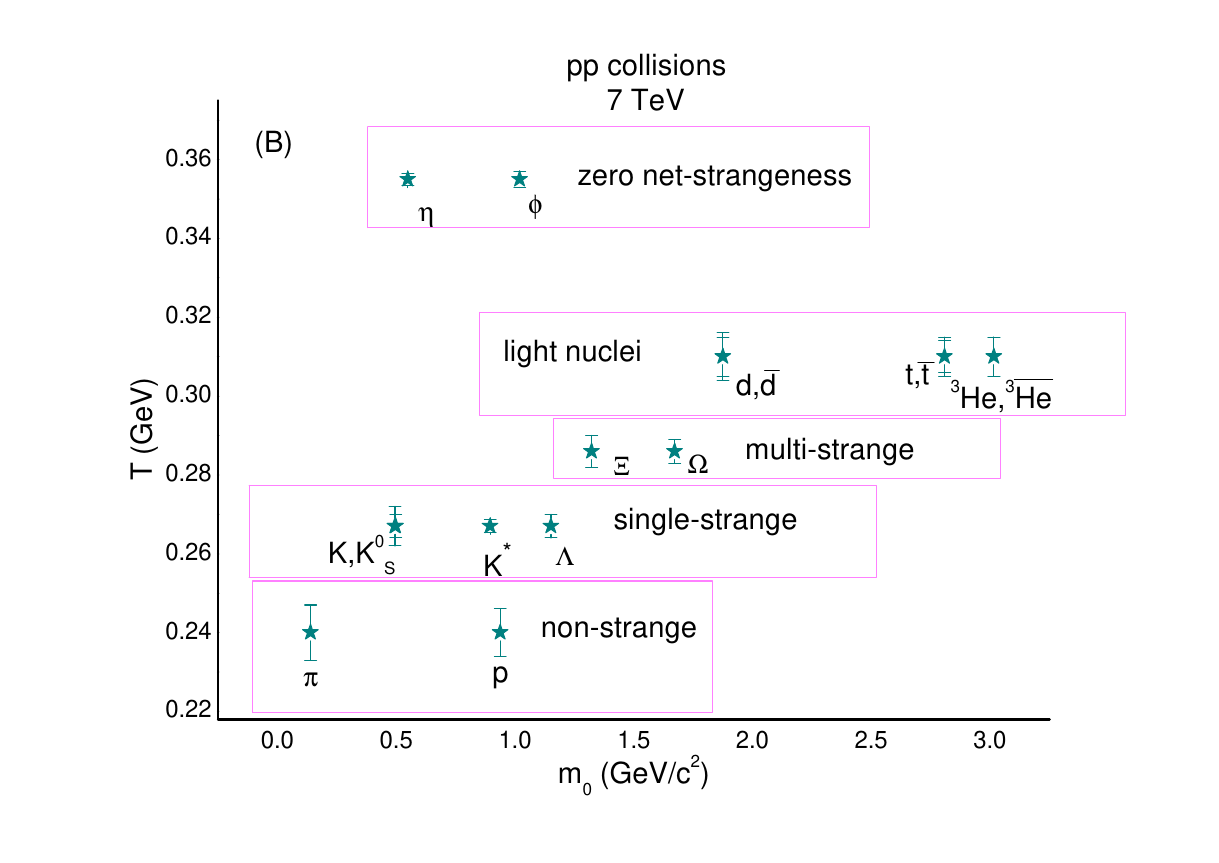}\hspace{-1.0cm}
\vspace*{-5mm}
\caption {Dependence of the effective temperature $T$ on the rest
  mass $m_{0}$ of a variety of particle species.} 
\label{fig3}
\end{figure*}

Figure \ref{fig3} shows the effective temperature \(T\) obtained
from fits of the TP-like function (\ref{eq4}) to the transverse
momentum spectra of identified particles in $pp$ collisions at
\(\sqrt{s}=2.76\) and \(7\)~TeV in panel (a) and (b), respectively.

At \(2.76\)~TeV, the analyzed species include \(\pi^++\pi^-\),
\(K^++K^-\), \(p+\bar p\), \(K^0_S\), \(\phi\), \(\Lambda\),
\(\bar\Lambda\), \(\Xi^-\), \(\bar\Xi^+\), \(\Omega^-+\bar\Omega^+\),
\(d\), and \(\bar d\). No $\eta$ meson data are available at
2.76~TeV in this analysis. The extracted \(T\) values do not
follow a simple mass ordering. The pions and protons share the
same lowest value, about \(0.200\pm0.004\)~GeV, even though their
masses differ by almost a factor of seven. The singly-strange kaons,
\(K^0_S\), \(\Lambda\), and \(\bar\Lambda\) are around
\(0.226\pm0.003\)~GeV. \(K^++K^-\) and \(K^0_S\) have approximately
the same mass, and their extracted \(T\) values are also essentially
the same. The multi-strange baryons \(\Xi\) and \(\Omega\) are slightly
higher, near \(0.240\pm0.003\)~GeV. The deuterons and anti-deuterons are
still higher, around \(0.257\pm0.003\)~GeV, and the \(\phi\) meson
is the highest, near \(0.311\pm0.006\)~GeV. The proton and \(\phi\)
have nearly the same mass, but their effective temperatures are very
different. So, while \(K^++K^-\) and \(K^0_S\) show that similar masses
can have similar \(T\) values, the proton and \(\phi\) show that nearly
equal masses can have very different \(T\) values. This clearly means
that mass alone cannot explain the pattern. 

At \(7\)~TeV, the same five-layer structure appears, but shifted
to higher temperatures. The non-strange \(\pi^\pm\) and \(p+\bar p\)
are around \(0.240\pm0.007\)~GeV. The singly-strange \(K^\pm\),
\(K^0_S\), \(\Lambda+\bar\Lambda\), and \(K^{*0}+\bar K^{*0}\) are
around \(0.267\pm0.005\)~GeV, the multi-strange \(\Xi^-+\bar\Xi^+\)
and \(\Omega^-+\bar\Omega^+\) are around \(0.286\pm0.004\)~GeV,
and the light nuclei \(d\), \(\bar d\), \(t\), \(\bar t\),
\(^{3}\mathrm{He}\), and \(^{3}\overline{\mathrm{He}}\) are around
\(0.310\pm0.006\)~GeV. We note that the $t$, $\bar t$,
\(^{3}\mathrm{He}\) and \(^{3}\overline{\mathrm{He}}\) spectra
contain too few points to constrain the four free parameters of
Eq.~(\ref{eq4}) independently, so for these four species the number
of degrees of freedom is not positive and the corresponding entries
in Table~\ref{tab1} are not independent determinations. The light-nuclei
layer is therefore established by the deuteron and anti-deuteron
fits, for which the number of degrees of freedom is positive at both
energies. The $\phi$ and $\eta$ mesons have the highest values,
around 0.355~GeV. The ordering is identical to that at \(2.76\)~TeV
for the layers that are present at both energies.

With the errors
given in Table~\ref{tab1}, every adjacent layer separation exceeds
\(3\sigma\) at both collision energies: about \(3.2\sigma\) for
the non-strange to singly-strange step, \(3.3\sigma\) for the
singly-strange to multi-strange step, \(4.0\sigma\) for the
multi-strange to light-nuclei step, and about \(8\sigma\) for
the light-nuclei to \(\phi\) step at \(2.76~\mathrm{TeV}\), with
comparable values at \(7~\mathrm{TeV}\). One might wonder whether
it is legitimate to call the top group a layer when, at
\(2.76~\mathrm{TeV}\), it contains only the \(\phi\). The answer
is yes, because the \(\phi\) value, \(0.311 \pm 0.006~\mathrm{GeV}\),
lies \(0.054 \pm 0.007~\mathrm{GeV}\) above the light-nuclei
layer at \(0.257 \pm 0.008~\mathrm{GeV}\). This is a separation
of about \(8\sigma\), far beyond what a statistical fluctuation
of the lower group would produce, so the \(\phi\) is not simply
a high-side outlier of the light nuclei but a genuinely higher
temperature. The same conclusion does not rest on a single
particle at \(7~\mathrm{TeV}\), where both \(\phi\) and \(\eta\)
occupy the top layer and each is separated from the light-nuclei
layer by at least \(7\sigma\).

Within the two $pp$ energies analyzed
here, this stability serves as a consistency check.
It indicates that, for a fixed collision system,
the effective-temperature ordering is governed primarily by the
species-dependent interaction with the medium rather than by the
collision energy. The present analysis is deliberately restricted
to $pp$ collisions at 2.76 and 7 TeV, which allows us to isolate
the energy dependence within a single collision system. Extending
this study to $p$-A and A-A collisions would provide valuable
insight into the system-size dependence of the effective-temperature
hierarchy, but it would require a dedicated analysis that accounts
for the different medium lifetimes, densities, and expansion histories. 

The effective temperature \(T\) extracted here includes the effect of
the collective flow, so it is systematically larger than the kinetic
freeze-out temperature \(T_{\mathrm{kin}}\) obtained from blast-wave
fits. The two quantities are therefore not directly comparable,
and a numerical comparison of the central values would be
misleading. What is meaningful is the ordering: the
effective-temperature hierarchy found here agrees with the
ordering of \(T_{\mathrm{kin}}\) reported in blast-wave analyses
Refs.\ \cite{Waqas:2021rmb,Badshah:2026ryg}. For the \(\phi\)
in particular, the blast-wave analysis of Pb-Pb data in
Ref.~\cite{Neeraj:2026snx} shows that the \(\phi\) does not share
the bulk \(\pi/K/p\) freeze-out surface, which is consistent with
the early decoupling of the \(\phi\) inferred here from \(T\).
A quantitative mapping between \(T\) and \(T_{\mathrm{kin}}\) would
require a blast-wave fit with species-dependent freeze-out and
is beyond the scope of this work.

Figure \ref{fig3} illustrates that non-strange hadrons
\(\pi\) and \(p\) sit at the bottom of the hierarchy. They
have the largest interaction cross-sections with the surrounding
hadronic matter, typically around \(15\) to \(40\)~mb~\cite{Neeraj:2026snx},
so they remain coupled to the medium the longest. Their spectra
are shaped by extensive rescattering and full participation in
the collective expansion, which lowers the extracted effective
temperature. The singly-strange hadrons \(K\), \(K_S^0\), \(K^{*0}\),
and \(\Lambda\) lie in the middle. With one unit of strangeness,
their cross-sections are smaller than those of pions and protons.
They therefore decouple earlier than the non-strange species but
later than particles with more strangeness. This intermediate
position is consistent with the triple kinetic freeze-out scenario
in our previous work, where singly-strange hadrons freeze out
between non-strange and multi-strange particles~\cite{Waqas:2021rmb}.
The multi-strange baryons \(\Xi\) and \(\Omega\) cross-sections
with the hadronic medium are much smaller. This weaker coupling
allows them to leave the medium sooner, preserving a higher \(T\).
This ordering is consistent with the standard freeze-out criterion
\cite{Heinz:2007in}, in which a species remains coupled to the
medium as long as its interaction rate exceeds the expansion rate,
so that species with smaller cross-sections decouple earlier and
retain a higher effective temperature \cite{Neeraj:2026snx}. 

Light nuclei (\(d\), \(t\), \(^{3}\mathrm{He}\)) appear next,
with higher \(T\) than the multi-strange baryons. These composite
objects form through coalescence of nucleons in the later stages
of the collision. Their large spatial size and weak binding energy
make them interact only weakly with the hadronic medium. Their
formation also requires nearby nucleons with correlated momenta,
which is possible only before the medium becomes too dilute. As a
result, they decouple early and show a higher \(T\) than the
multi-strange baryons. This behavior has been studied previously,
where light nuclei \(T\) are found to be higher than those of bulk
hadrons~\cite{Waqas:2022fnl,Badshah:2026ryg}.

Finally, the
\(\phi(1020)\) meson, and at \(7\ \text{TeV}\) also the \(\eta\)
meson, has the highest value for \(T\) of all analyzed species.
Although the mass of \(\phi(1020)\) meson is close to the proton
mass, it does not follow the mass hierarchy. The reason is the
OZI (Okubo--Zweig--Iizuka) rule
\cite{Okubo:1963fa,Zweig:1964ruk,Iizuka:1966fk}. The \(\phi\)
is almost a pure \(s\bar{s}\) state with hidden strangeness,
and its interaction with non-strange hadrons is strongly
suppressed. Its cross-section with the hadronic medium is
therefore unusually small, giving it a long mean free path
and a short re-scattering depth. It decouples almost immediately
after hadronization. A recent Boltzmann--Gibbs blast-wave contour
analysis of ALICE Pb-Pb data has shown that the bulk \(\pi/K/p\)
freeze-out point is excluded at the \(4.1\sigma\) level
(\(\Delta\chi^2 = 21.7\)) from the \(\phi\) freeze-out parameters,
which supports this differential behavior~\cite{Neeraj:2026snx}.
The \(\phi\) therefore probes the earliest and hottest stage of
the system, before significant cooling and collective expansion.

Naturally, these layers establish a quintuple effective-temperature
hierarchy in which particles decouple sequentially according to
their strangeness content, interaction cross-section, and compositeness.
It is analogous to the quadruple freeze-out picture discussed in
Ref.\ \cite{Badshah:2026ryg}. The same five-layer pattern appears
at both \(2.76\) and \(7\)~TeV. The values of \(T\) increase
with the collision energy, while the ordering and grouping remain
stable. That stability makes the quintuple effective-temperature
hierarchy a robust feature of the data.

We would like to highlight
that, in the present work, at $7$~TeV the $\eta$ meson is extracted
with a temperature that is statistically indistinguishable from
that of the $\phi$. The two values are $T_\phi = 0.355 \pm 0.002$~GeV
and $T_\eta = 0.355 \pm 0.0016$~GeV, so their difference is
$0.000 \pm 0.0026$~GeV and they agree to within about $5$~MeV
at $95\%$ confidence. For the $\phi$, the reason for its high
temperature is well understood, as discussed above. The $\eta$,
however, is not OZI-protected; the axial anomaly strongly violates
the OZI rule in the pseudoscalar sector, and its $\eta$-nucleon
cross-section is not small. So why does it sit at the same high
effective temperature as the $\phi$? We do not have a clear answer.

A similar tension has been noted in thermal-model analyses at SIS
energies, where the $\eta$ freeze-out line was found not to cross
the common chemical freeze-out curve extracted from other particle
species ~\cite{Cleymans:1998yb}, though the physical origin of this
discrepancy remains an open question. Our observation adds a new
aspect to this puzzle: the $\eta$ and $\phi$, despite very
different interaction properties, are found at the same effective
temperature within the present precision. We note that the $\eta$
fit has $\chi^2/\mathrm{dof} = 31/9 \approx 3.4$, which is
acceptable, but not as good as
the $\phi$ fit ($12/17 \approx 0.71$); if the $\eta$ uncertainty
is scaled by the fit quality, the two temperatures still agree
within about $7$~MeV at $95\%$ confidence. This coincidence may
point to a common early production mechanism or to in-medium
effects that we do not yet fully understand. A fully quantitative
description of the sequential decoupling, including the $\eta$
puzzle, would require a dedicated transport calculation or a
blast-wave analysis with species-dependent freeze-out, which is
beyond the scope of the present work.

\begin{table*}
\begin{center}
\begin{tabular}{ccccccccc}\\ \hline\hline
Energy           & Particle    & $T$ (GeV)        & $n_0$  & $a_0$
& $\chi^2$/ dof \\ \hline
         2.76 TeV        &  $\pi^++\pi^-$                  & $0.200\pm0.004$    & $6.0\pm0.5$          & $0.05\pm0.004$     & 148/59\\
                         &  $K^++K^-$                      & $0.226\pm0.003$    & $6.1\pm0.7$          & $0.16\pm0.002$     & 132/59\\
                         &  $p+\bar p$                     & $0.200\pm0.007$    & $6.25\pm0.4$         & $0.164\pm0.004$    & 117/34\\
                         &  $K^0_S$                        & $0.226\pm0.003$    & $6.2\pm0.5$          & $0.008\pm0.0004$   & 39/10\\
                         &  $\Phi$                         & $0.311\pm0.006$    & $7.8\pm0.8$          & $0.650\pm0.008$    & 6/3\\
                         &  $\Lambda$                      & $0.226\pm0.003$    & $8.0\pm0.6$          & $0.200\pm0.006$    & 89/29\\
                         &  $\bar \Lambda$                 & $0.226\pm0.003$    & $8.0\pm0.5$          & $0.200\pm0.006$    & 104/31\\
                         &  $\Xi^-$                        & $0.240\pm0.003$    & $8.0\pm0.4$          & $0.235\pm0.007$    & 27/9\\
                         &  $\bar \Xi^+$                   & $0.240\pm0.003$    & $8.2\pm0.4$          & $0.235\pm0.006$    & 23/10\\
                         &  $\Omega^-+\bar \Omega^+$       & $0.240\pm0.003$    & $8.5\pm0.6$          & $0.430\pm0.011$    & 5/3\\
                         &  $d$                            & $0.257\pm0.003$    & $18\pm1.2$           & $0.100\pm0.0006$   & 7/3\\
                         &  $\bar d$                       & $0.257\pm0.002$    & $18\pm1.2$           & $0.100\pm0.0006$   & 7/3\\
\hline
  7 TeV                  &  $\pi^{\pm}$                    & $0.240\pm0.007$    & $6.2\pm0.3$          & $0.070\pm0.006$    & 31/54\\
                         &  $K^{\pm}$                      & $0.267\pm0.005$    & $6.4\pm0.2$          & $0.258\pm0.008$    & 92/42\\
                         &  $p+\bar p$                     & $0.240\pm0.006$    & $7\pm0.3$            & $0.280\pm0.010$    & 139/45\\
                         &  $K^0_S$                        & $0.267\pm0.003$    & $6.2\pm0.4$          & $0.220\pm0.012$    & 137/40\\
                         &  $\Phi$                         & $0.355\pm0.002$    & $8.2\pm0.8$          & $0.520\pm0.011$    & 12/17\\
                         &  $\Lambda+\bar \Lambda$         & $0.267\pm0.003$    & $7.2\pm0.5$          & $0.300\pm0.009$    & 112/29\\
                         &  $K^{*0}+\bar K^{*0}$           & $0.267\pm0.005$    & $7\pm0.5$            & $0.033\pm0.007$    & 39/24\\
                         &  $\Xi^-+\bar \Xi^+$             & $0.286\pm0.004$    & $8\pm1.2$            & $0.020\pm0.002$    & 38/14\\
                         &  $\Omega^-+\bar \Omega^+$       & $0.286\pm0.003$    & $8\pm0.4$            & $0.286\pm0.012$    & 2.8/4\\
                         &  $\eta$                         & $0.355\pm0.0016$    & $8.2\pm1.1$          & $0.400\pm0.012$    & 31/9\\
                         &  $d$                            & $0.310\pm0.005$    & $29\pm2.4$           & $0.060\pm0.0008$  & 40/17\\
                         &  $\bar d$                       & $0.310\pm0.006$    & $32\pm2.4$           & $0.060\pm0.0004$   & 29/16\\
                         &  $t$                            & $0.310\pm0.005$    & $25\pm2.4$           & $0.010\pm0.0003$    & 0.7/--\\
                         &  $\bar t$                       & $0.310\pm0.004$    & $25\pm2.3$           & $0.010\pm0.0004$   & 0.5/--\\
                         &  ${}^{3}\mathrm{He}$            & $0.310\pm0.005$    & $25\pm2.1$           & $0.007\pm0.0005$   & 0.3/--\\
                         &  ${}^3\overline{\mathrm{He}}$   & $0.310\pm0.005$    & $25\pm2.2$           & $0.010\pm0.0003$   & 3.6/--\\
\hline
\end{tabular}
\end{center}
\caption{Values of the fitting parameters:
the effective temperature $T$, the non-extensivity parameter
$n_0$, the exponent $a_0$, $\chi^2$ and number of degrees of freedom (dof)
corresponding to the fits in Figures \ref{fig1-b} and \ref{fig2-b}.
For \(t\), \(\bar{t}\), \(^{3}\mathrm{He}\) and
\(^{3}\overline{\mathrm{He}}\), the published spectra contain
fewer points than the four free parameters of Eq.~(\ref{eq4}), so
the number of degrees of freedom is not positive;
their entries are listed for completeness only and are not used
in the analysis of the effective-temperature hierarchy.}
\label{tab1}
\end{table*}

\begin{figure*}
\centering
\hspace*{-10mm}
\includegraphics[width=0.55\textwidth]{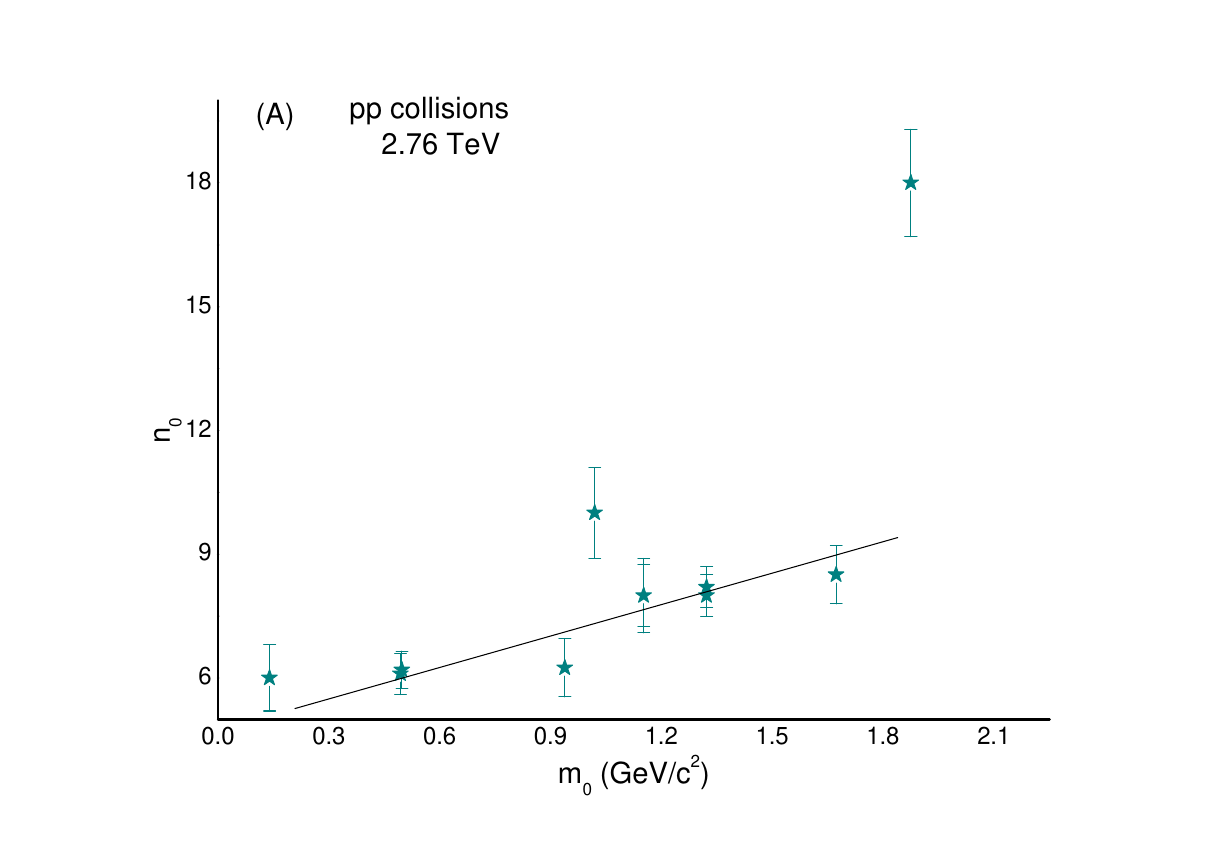}\hspace{-1.5cm}
\includegraphics[width=0.55\textwidth]{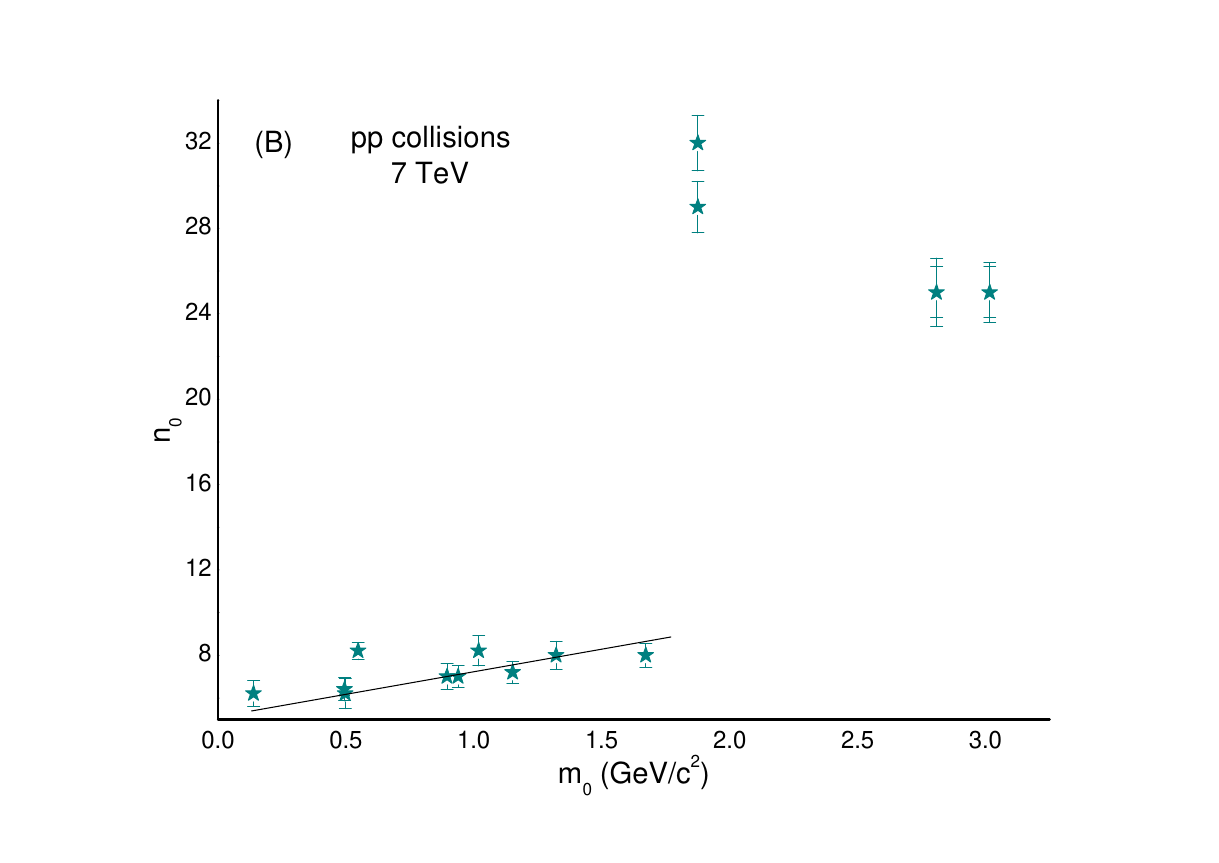}\hspace{-1.0cm}
\vspace*{-5mm}
\caption {Mass dependence of the non-extensivity parameter $n_0$
  in Eq.\ (\ref{eq4}) for multiple particle species; they can be
  identified in Figure \ref{fig3}. According to
  Eq.\ (\ref{eq2}) it fixes the parameter $q = 1 + 1/n_{0}$ in the
  $q$-exponential function of the Tsallis distribution; the limit
  $q \to 1$ corresponds to the standard exponential of the Boltzmann
  distribution.} 
\label{fig4}
\end{figure*}

Figure~\ref{fig4} is similar to Figure~\ref{fig3}, but it shows
the extracted \(n_0\) values for individual species at the
lower and higher collision energy. At the lower energy, pions
have the lowest \(n_0\), and therefore the maximal value of the
$q$-parameter, which quantifies the non-extensivity.
Kaons lie just above them, and the
proton is a little higher. The \(\Lambda\) and \(\bar{\Lambda}\)
move up, the \(\Xi\) and \(\bar{\Xi}\) are in the same region,
and \(\Omega^-+\bar{\Omega}^+\) is the highest among the hadrons.
The \(\phi\) meson sits with the heavier-hadron group, while
deuterons and anti-deuterons are much larger. At the higher
energy the values are generally larger, but the grouping is
similar: pions remain lowest, kaons are close by, the proton
is above them, and the \(\Lambda/\bar{\Lambda}\) and
\(K^{*0}/\bar{K}^{*0}\) form the next group, with \(\Xi/\bar{\Xi}\)
and \(\Omega/\bar{\Omega}\) higher; \(\phi\) and \(\eta\) lie
near the \(\Omega\) group. The light nuclei are far above the
hadrons, with \(d\) and \(\bar{d}\) at the top and \(t\),
\(\bar{t}\), \({}^{3}\mathrm{He}\), and \({}^{3}\overline{\mathrm{He}}\)
clustering at the lower end of that nuclear group.

The non-strange hadrons occupy the lowest range of \(n_0\). These
species have large hadronic cross-sections and remain coupled to
the medium until late times, so their distributions carry the
accumulated effect of re-scattering and depart most from
equilibrium. The singly-strange kaons remain close to the
non-strange hadrons, while the \(\Lambda\), \(\Xi\), and
\(\Omega\) give larger values. The steady increase from the
non-strange sector through the \(\Lambda\) to the multi-strange
baryons shows that the distribution moves closer to equilibrium
as the interaction cross-section with the medium decreases.
Particles that decouple earlier have less time to develop a
non-equilibrium distribution through re-scattering.

The \(\phi(1020)\) meson does not follow the same pattern as
the hadrons discussed above. Its \(n_0\) value is higher than
that of the proton and the kaons, even though its mass is
close to the proton mass.
Due to the OZI suppression, the \(\phi\) decouples early and its
distribution is less affected by re-scattering than those of
the non-strange hadrons. A similar conclusion has been reached
in Pb-Pb collisions at 2.76~TeV, where a Boltzmann-Gibbs
blast-wave contour analysis shows that the \(\phi\) does not
share the bulk \(\pi/K/p\) freeze-out surface~\cite{Neeraj:2026snx}.
This confirms that the interaction strength with the medium, not the
rest mass alone, plays a decisive role in shaping the final distribution. 

The ordering in \(n_0\) should be understood as an ordering
in the non-extensivity, rather than a kinetic
freeze-out hierarchy. The \(n_0\) values give a measure of
spectral equilibration and are separate from the effective-temperature
hierarchy and the kinetic freeze-out sequence derived from
blast-wave analyses.

\begin{figure*}
\centering
\hspace*{-5mm}  
\includegraphics[width=0.55\textwidth]{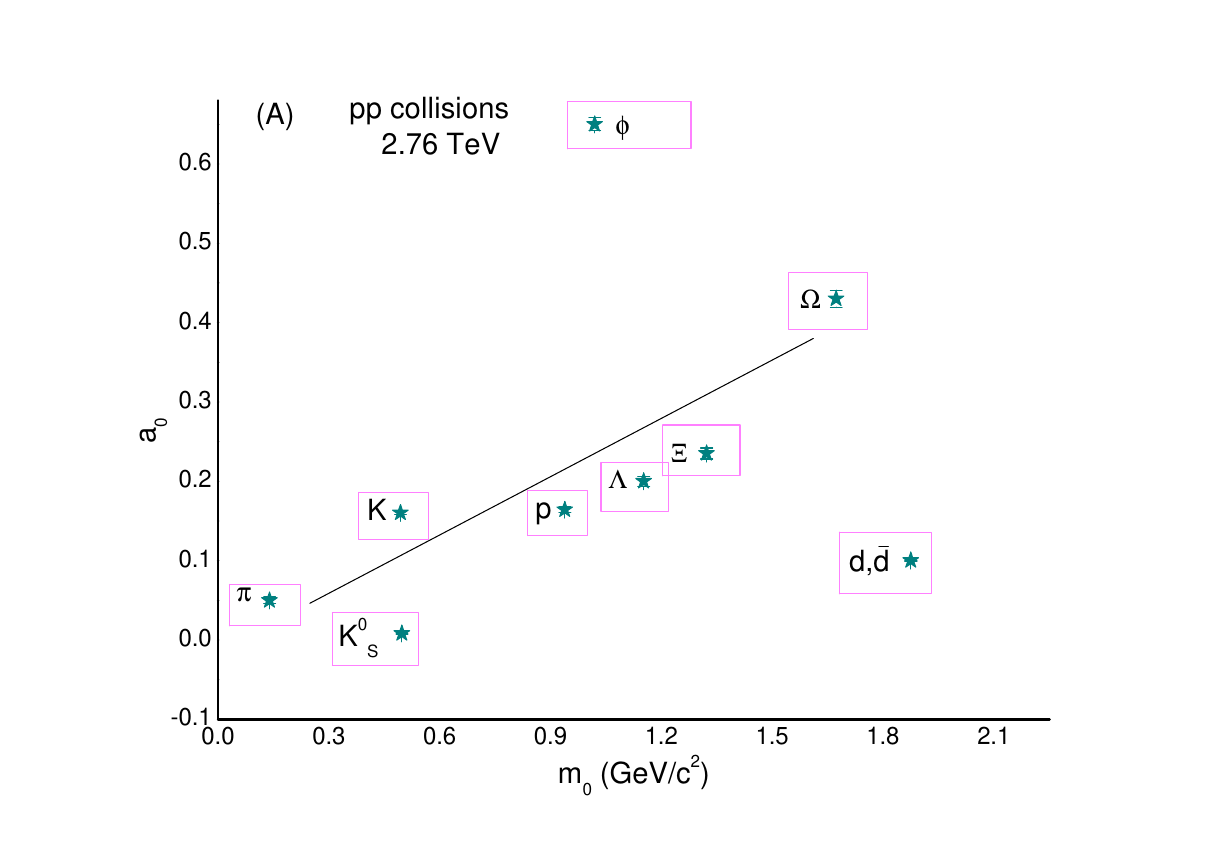}\hspace{-1.5cm}
\includegraphics[width=0.55\textwidth]{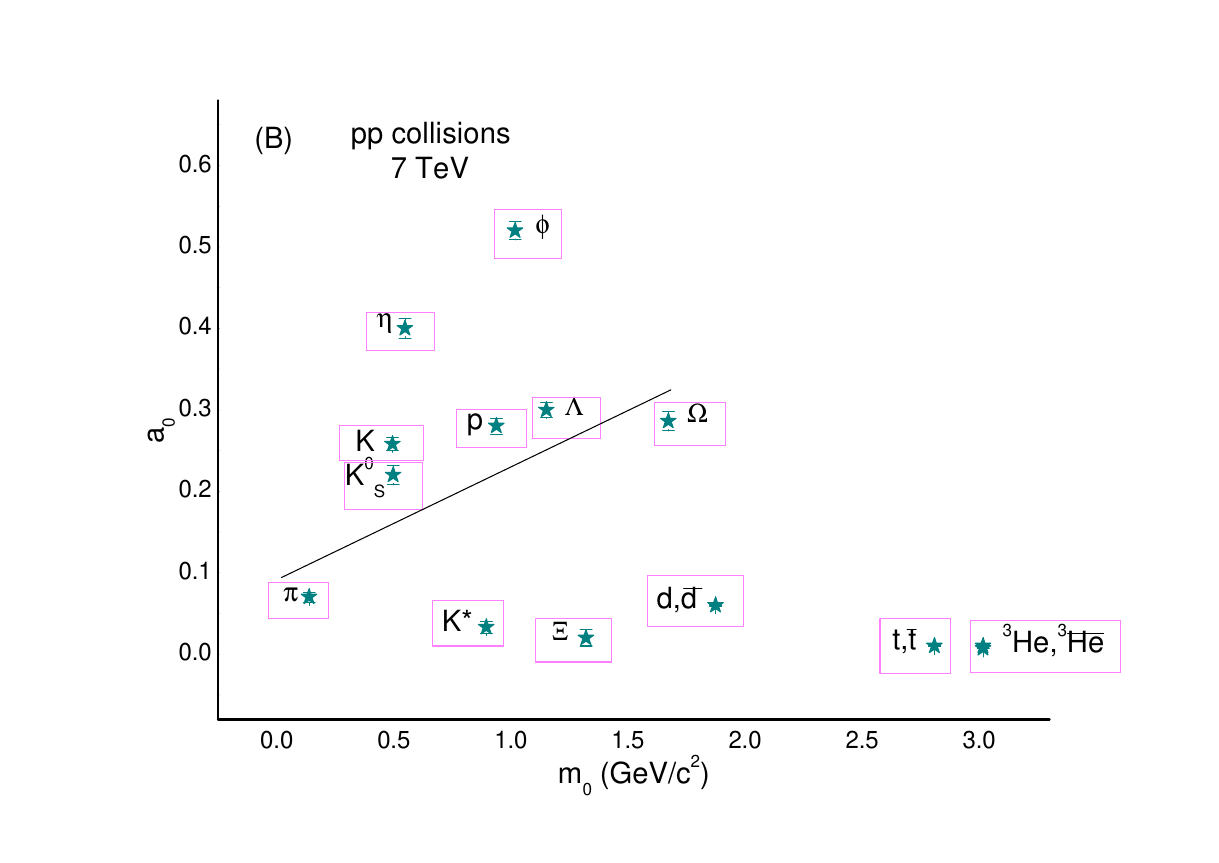}\hspace{-1.0cm}
\vspace*{-5mm}
\caption {This figure exhibits the dependence of the exponent
  $a_0$ on particle rest mass.} 
\label{fig5}
\end{figure*}

The light nuclei are produced mainly by coalescence of nucleons
near the end of the collision. Their production mechanism is
distinct from that of individual hadrons, and their \(n_0\)
values reflect this difference. The very large \(n_0\) indicates
that their distribution is much closer to the equilibrium
than that of any hadron. However, because their production is
dominated by coalescence, this should not be interpreted as
full thermal equilibration. The large \(n_0\) may instead
reflect the coalescence mechanism and the late-stage correlation
of nucleons.

The one case that does not follow the hadronic
pattern is the \(\eta\) meson at \(7\) TeV. Unlike the \(\phi\),
the \(\eta\) is not OZI-protected, and its interaction with
nucleons is not suppressed.  Its \(n_0\) value ($8.2 \pm 1.1$)
is consistent with that of the \(\phi\) ($8.2 \pm 0.8$).
We have no explanation for this within the
present analysis and record it as an open question.

The non-extensivity parameter \(n_0\) follows a related but not
identical hierarchy to the effective temperature \(T\).
Non-strange hadrons give the smallest values, singly- and
multi-strange hadrons are intermediate, and light nuclei give
the largest values. The \(\phi\) and \(\eta\) mesons do not
follow the \(T\)-ordering; their \(n_0\) values are comparable
to the multi-strange baryons, not the highest in the set. This
indicates that \(n_0\) is sensitive to both early decoupling
and the specific production mechanism, especially coalescence
for light nuclei.

Figure \ref{fig5} presents the result for \(a_0\). As discussed in
Section~\ref{method}, the parameter \(a_0\) is the correction index
introduced in the TP-like function to improve the description
of the \(p_T\) distribution. Following the interpretation in
Ref.~\cite{Yang:2020wof}, it describes the bending degree of
the spectrum at very low \(p_T\). A negative \(a_0\) reflects
an increment in the yield of light particles through resonance
decay, while a positive \(a_0\) reflects a decrement in the
yield of heavy particles due to decay or absorption in the hot
and dense medium. It should be noted, however, that this
interpretation was developed in a quark-level analysis, so its
direct application to a hadron-level analysis, such as the present
one, should be made with caution. 

At \(2.76\)~TeV, the \(a_0\) values span a fairly wide
range. The smallest is found for \(K^0_S\), while the largest
belongs to the \(\phi\) meson. The pions, kaons, protons,
\(\Lambda\), \(\Xi\), \(\Omega\), and deuterons all fall in
between, with a mild upward trend from lighter to heavier species.
At \(7\)~TeV the values are generally larger, but this trend is
not uniform. For instance, \(K^0_S\) moves up considerably compared
with \(2.76\)~TeV, and the \(\phi\) remains on top, while
a few species such as \(K^{*0}\) and \(\Xi\) sit at the low end.
Overall, the energy dependence is not a simple rescaling; some
particles shift strongly, while others hardly change.

For the species considered here, the magnitude of the extracted
\(a_0\) does not follow a simple mass ordering, nor does it
reproduce the same hierarchy as \(T\) or \(n_0\).
At \(2.76\) TeV, \(K^0_S\) has the smallest
\(a_0\) in the entire set, whereas the \(\phi\) meson has the
largest, even though its mass is close to the proton mass. At
\(7\) TeV, the \(\Xi^- + \bar{\Xi}^+\) gives a very small value
(\(0.020\pm0.002\)), whereas the \(\eta\) meson, which has a mass
of only \(0.548\)~GeV, gives \(0.400\pm0.012\). The light nuclei
give the smallest values at this energy, between \(0.007\) and
\(0.060\), despite being the heaviest species in the analysis.
These observations indicate that \(a_0\) is sensitive to the
specific production and interaction properties of each species,
and that its physical interpretation in the present hadron-level
analysis is not as straightforward as in the participant quark-level
analysis of Ref.~\cite{Yang:2020wof}. For this reason, we treat \(a_0\)
as a phenomenological fit parameter rather than a direct probe of
the effective-temperature hierarchy.

\begin{figure*}
\vspace*{-5mm}
\centering
\hspace*{-5mm}  
\includegraphics[width=0.55\textwidth]{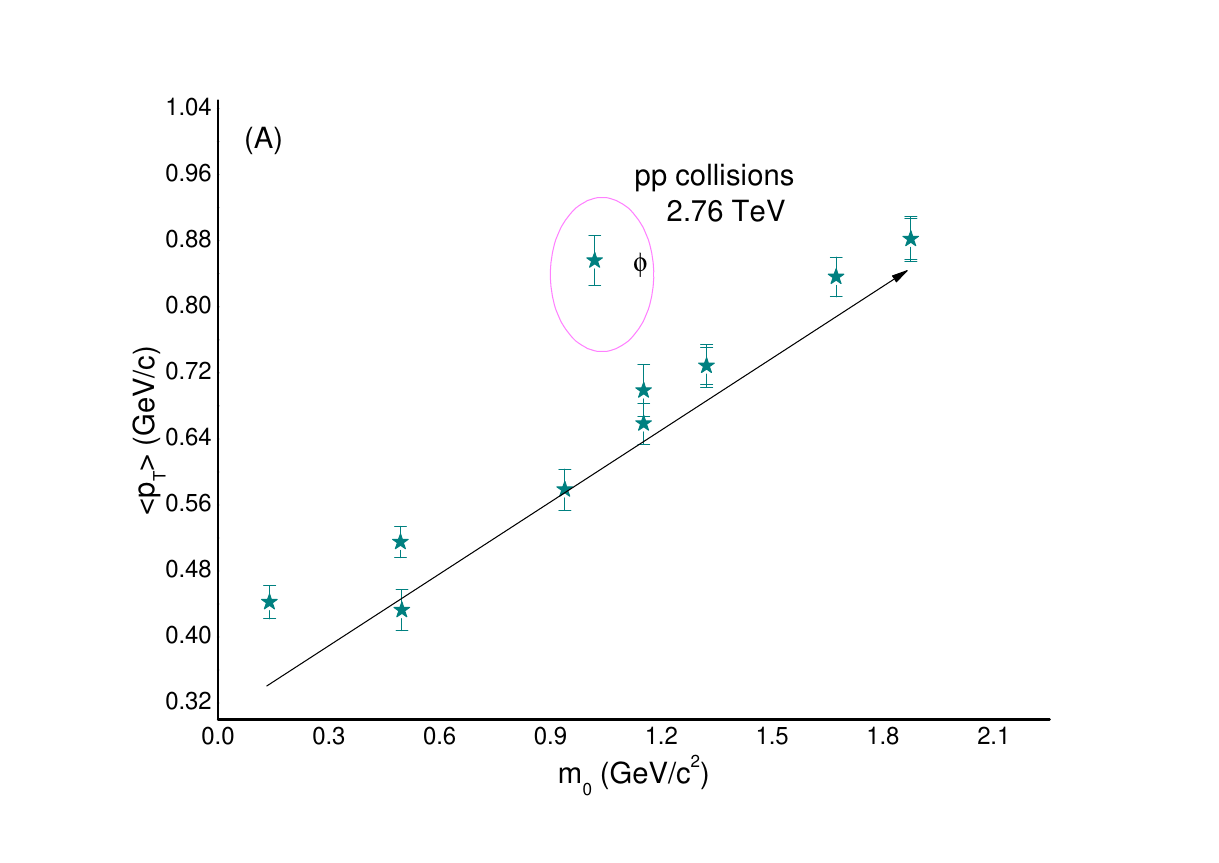}\hspace{-1.5cm}
\includegraphics[width=0.55\textwidth]{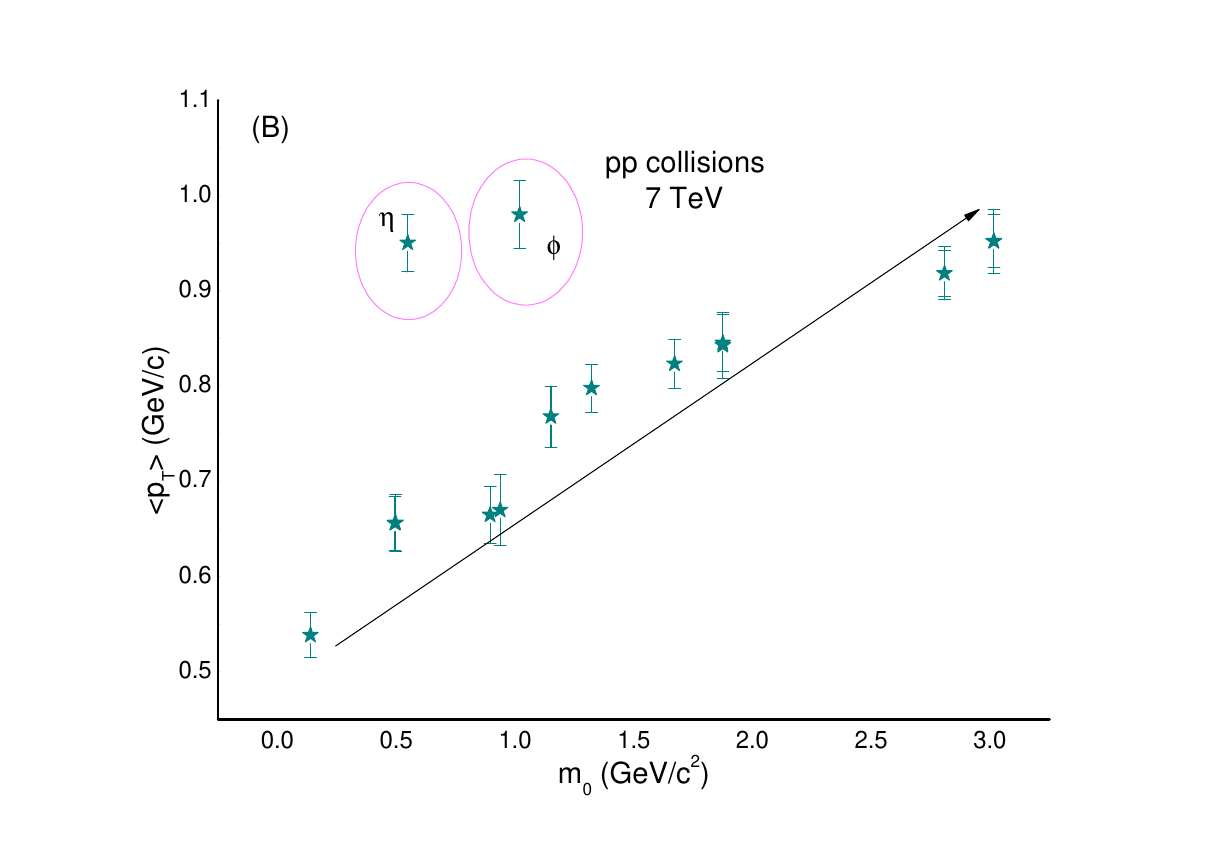}\hspace{-1.0cm}
\vspace*{-5mm}
\caption {Dependence of $\la p_T \ra$ on the particle rest mass $m_{0}$.} 
\label{fig6}
\end{figure*}

Figure~\ref{fig6} displays the mean transverse momentum
\(\langle p_T \rangle\) as a function of particle species
at \(2.76\) and \(7~\mathrm{TeV}\) in panels (A) and (B).
At \(2.76~\mathrm{TeV}\), \(\langle p_T \rangle \) increases from
pions to kaons to protons, while the \(K_S^0\) falls below
the pion value even though the kaon mass is larger. The \(\phi\)
meson, the \(\Lambda\) and \(\bar{\Lambda}\), the \(\Xi^-\) and
\(\bar{\Xi}^+\), the \(\Omega^- + \bar{\Omega}^+\), and the
$d$ and $\bar d$ appear at higher values, with the
$d$ and $\bar d$ giving the largest \(\langle p_T \rangle \)
among the species shown. On the other hand, at \(7~\mathrm{TeV}\),
\(\langle p_T \rangle \) is higher than at \(2.76~\mathrm{TeV}\)
for every species. The pions have the lowest values, followed by
the kaons and protons, while the \(K_S^0\) and \(K^{*0} + \bar{K}^{*0}\)
are comparable. The \(\eta\) and \(\phi\) mesons,
together with the \(\Lambda + \bar{\Lambda}\), \(\Xi^- + \bar{\Xi}^+\),
and \(\Omega^- + \bar{\Omega}^+\), show progressively larger
\(\langle p_T \rangle \). The light nuclei, including $d$, $\bar d$
$t$, $\bar t$, \(^3\mathrm{He}\), and \(^3\overline{\mathrm{He}}\),
exhibit the highest \(\langle p_T \rangle \) among the species considered.
The \(\phi\) meson gives a larger value than the proton at both
energies. The \(\eta\) and \(\phi\) mesons have the largest values
among the hadrons at \(7\ \mathrm{TeV}\). Note that these values are
extracted within the framework of the TP-like function and are therefore
model-dependent.

\begin{figure*}
\vspace*{-8mm}
\centering
\hspace*{-5mm}
\includegraphics[width=0.55\textwidth]{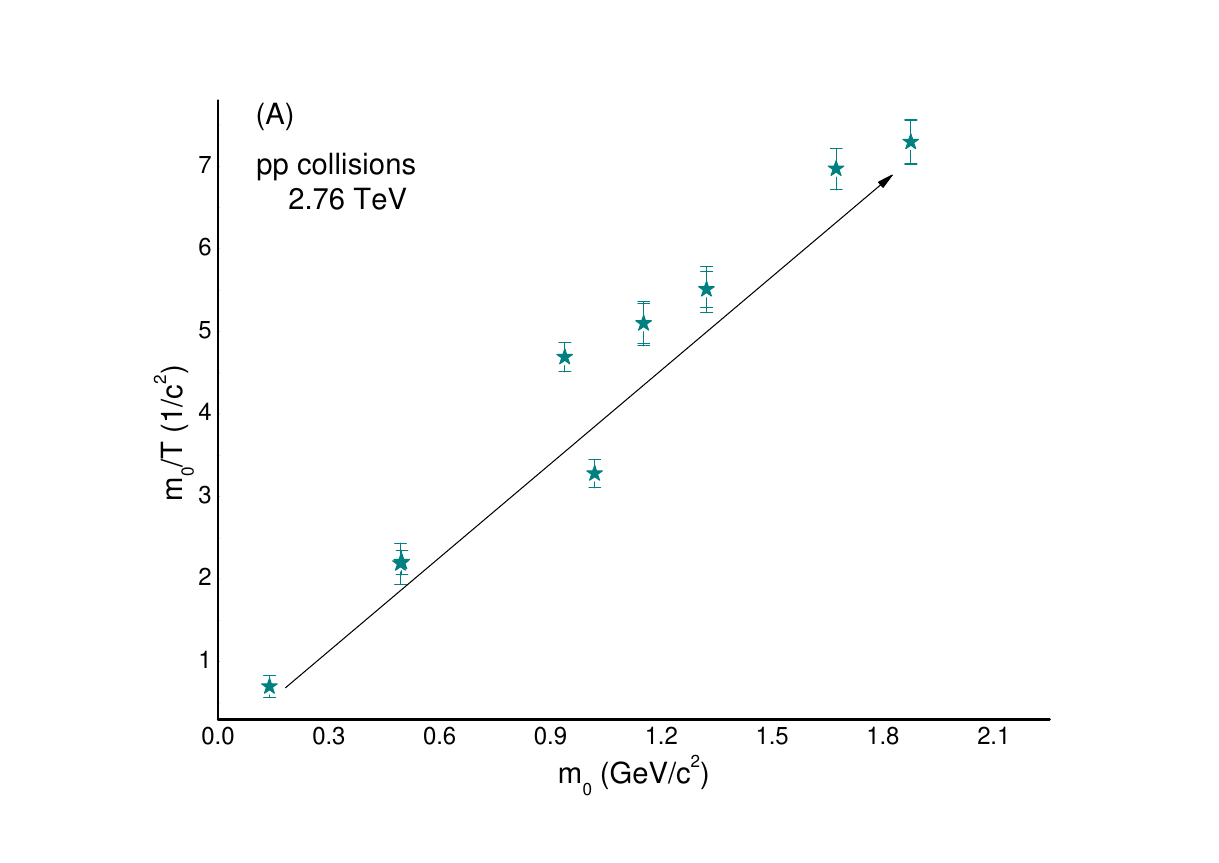}\hspace{-1.5cm}
\includegraphics[width=0.55\textwidth]{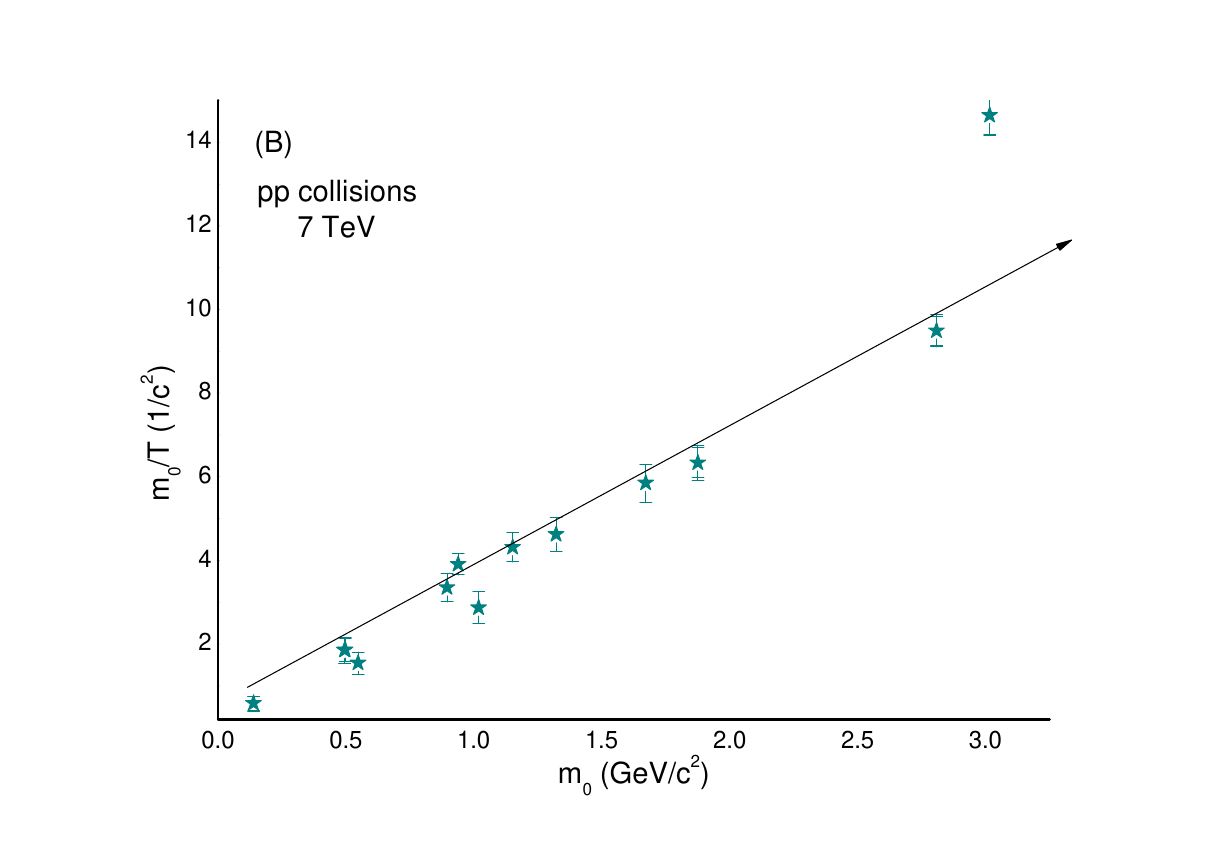}\hspace{-1.0cm}
\vspace*{-5mm}
\caption {Dependence of the ratio $m_0/T$ on the particle rest mass $m_{0}$.} 
\label{fig7}
\end{figure*}

Figure~\ref{fig7} shows the ratio of the particle rest mass to the
effective temperature, \(m_0/T\), as a function of particle species
at \(2.76\) and \(7~\mathrm{TeV}\) in panels (A) and (B), respectively.
At \(2.76~\mathrm{TeV}\), the pions give the smallest ratio. The kaon
ratios are close to each other and larger than the pion ratios,
while the proton ratio is even higher. The \(\phi\) meson lies between
the kaon and proton values. The \(\Lambda\) and \(\bar{\Lambda}\)
come next, followed by the \(\Xi^-\) and \(\bar{\Xi}^+\), and then
the \(\Omega^- + \bar{\Omega}^+\). The deuterons and anti-deuterons
give the largest ratio among the species shown.

At the higher collision energy of 7~TeV, the pions again give the smallest
\(m_0/T\) ratio. The \(\eta\) meson lies above the pions but below the
kaons, while the $K^{\pm}$ and \(K^0_S\) are close to each other.
The \(\phi\) meson comes next, followed by the \(K^{*0} + \bar{K}^{*0}\)
and the proton, with the proton slightly higher than the
\(K^{*0} + \bar{K}^{*0}\). The \(\Lambda + \bar{\Lambda}\),
\(\Xi^- + \bar{\Xi}^+\), and \(\Omega^- + \bar{\Omega}^+\) show
a steady increase. The light nuclei give the largest ratios,
rising from the $d$ and $\bar d$ to the $t$ and $\bar t$
and then to \(^{3}\mathrm{He}\) and \(^{3}\overline{\mathrm{He}}\). 

The ratio \(m_0/T\) measures how the rest mass of a particle compares
to the thermal energy scale at freeze-out. In a thermal system, the
equilibrium abundance of a species is proportional to \(e^{-m_0/T}\).
When \(m_0/T\) is small, the exponential is close to 1 and the
species is abundant. When \(m_0/T\) is large, the exponential
becomes very small and the species is suppressed. This is Boltzmann
suppression, which is why lighter particles have larger thermal
abundances. In the present analysis, the pions have the smallest
\(m_0/T\), followed by the kaons. The protons and \(\phi\) meson
lie at intermediate values above the kaons. The multi-strange
baryons are found at larger \(m_0/T\), and the light nuclei give
the largest values among the species considered. For the light
nuclei, the Boltzmann factor \(e^{-m_0/T}\) is very small, so a
thermal population at freeze-out would be negligible. This is
consistent with the coalescence mechanism for light nuclei
production, which has been established in previous studies
\cite{ALICE:2017xrp,Neidig:2021bal,Wang:2023rpd,Wang:2024hok,Wang:2017vsm}.

The ratio  \(m_0/T\) increases with particle mass at both energies. The
values at \(7\)~TeV are smaller than those at \(2.76\) TeV for the same
species because the effective temperature is larger at the higher
energy. Two exceptions to the simple mass trend appear in the data.
At \(2.76\) TeV, the \(\phi\) meson has a smaller ratio \(m_0/T\) than
the proton even though its mass is close to the proton mass. At
\(7\) TeV, the \(\eta\) meson has a smaller ratio \(m_0/T\) than the kaon
even though its mass is close to the kaon mass. In both cases, the
smaller ratio reflects the larger effective temperature extracted
for these particles, which follows from their early decoupling. The
light nuclei show the opposite behavior: their \(m_0/T\) ratios are
the largest in the set because their masses are large and their
effective temperatures are only moderately higher than those of
the hadrons. This pattern is consistent with the
effective-temperature hierarchy established from \(T\) and \(n_0\).

\vspace{-2mm}
\section{Conclusions}
\label{conclu}

We have analyzed the transverse momentum spectra of a broad set
of identified particles produced in proton-proton collisions at
$\sqrt{s}=2.76$ and $7$~TeV, using the modified Tsallis-Pareto-type
function to extract the effective temperature $T$, the
non-extensivity parameter $n_0$, and the correction index $a_0$.
The goal was to test whether the effective-temperature hierarchy
extends beyond the recently established quadruple scenario.

Our results give a clear answer. The effective temperature organizes
the particle species
into five distinct layers rather than four, establishing a quintuple
effective-temperature hierarchy. The non-strange hadrons occupy
the lowest layer, followed sequentially by singly-strange hadrons,
multi-strange baryons, light nuclei, and finally the
\(\phi(1020)\) meson, joined at \(7~\mathrm{TeV}\) by the \(\eta\)
meson, at the top. At \(2.76\ \text{TeV}\), the top layer is the
\(\phi(1020)\) meson. At \(7\ \text{TeV}\), the \(\phi\) and \(\eta\)
mesons populate the highest-temperature layer, with temperatures
that agree within about 5 MeV at 95\% confidence. With the errors
given in Table~\ref{tab1}, every adjacent layer separation exceeds
\(3\sigma\) at both collision energies, and the top separation exceeds
\(7\sigma\). The ordering and grouping remain stable at both
collision energies, while only the absolute values of \(T\) shift
upward. This stability shows that the ordering in the effective
temperature is governed by how each species interacts with the
medium, not by the collision energy. 

Mass alone cannot explain the pattern. The proton and the \(\phi\)
meson have nearly equal masses but very different effective
temperatures. The \(\phi\), being an almost pure \(s\bar{s}\) state,
has an OZI-suppressed coupling to non-strange hadrons, decouples
early, and retains a high temperature. The non-strange hadrons
have the largest cross-sections, stay coupled longest, and have
the lowest temperatures. Singly- and multi-strange hadrons fall
in between, following their interaction strengths. Light nuclei
sit above the multi-strange baryons because their weak binding
and large size make them interact only weakly with the medium,
and their coalescence formation requires correlated nucleons
before the medium becomes too dilute.

The non-extensivity parameter \(n_0\) follows a pattern that is
similar to, but not identical with, that of the effective
temperature \(T\). Non-strange hadrons have the smallest \(n_0\)
values, which means that they depart furthest from equilibrium. Light
nuclei have the largest values, indicating spectra closest to
equilibrium. The \(\phi\) and \(\eta\) mesons, however, do not
follow the \(T\)-ordering; their \(n_0\) values are comparable
to those of the multi-strange baryons. This suggests that \(n_0\)
is sensitive both to early decoupling and to the specific production
mechanism, especially coalescence in the case of light nuclei.

The correction index \(a_0\), by contrast, does not follow this
hierarchy and appears to depend on the particular production and
interaction properties of each species. We therefore treat it
as a phenomenological fit parameter rather than a direct probe
of the effective-temperature hierarchy. The ratio \(m_0/T\)
increases with mass, as expected, with the \(\phi\) and \(\eta\)
as exceptions that reflect their early decoupling. The large
values for light nuclei support a coalescence origin rather than
a thermal population.

One result remains unexplained. At $7$~TeV, the \(\eta\) meson is
found at a temperature that agrees with the \(\phi\) to within
about $5$~MeV at $95\%$ confidence, and its $n_0$ value ($8.2 \pm 1.1$)
is also consistent with that of the $\phi$ ($8.2 \pm 0.8$), even
though the $\eta$ is not OZI-protected and its interaction with
nucleons is not suppressed. There is no obvious reason why the
$\eta$ should decouple as early as the $\phi$. This could point
to a common early production mechanism or to in-medium effects that
are not yet understood, and we record it as an open question.

\noindent
{\bf Data availability}
The data supporting the findings of this study are provided in the
article and are referenced appropriately throughout the text.\\

\noindent
{\bf Compliance with Ethical Standards}
The authors confirm that this paper complies with all applicable
ethical standards.\\

\noindent
{\bf Acknowledgments:}
This work was supported by Hubei University of Automotive Technology,
Shiyan, China under the grant number BK202313, and  by the Mexican
funding agency UNAM-DGAPA through project PAPIIT IG100826 ``Estudios
de diagrama de fase con m\'{e}todos experimentales y de Monte Carlo''.
\\
\end{multicols}
\end{document}